\documentclass[useAMS,usenatbib]{mn2e}
\usepackage{graphicx}
\usepackage{amsfonts}
\usepackage{pifont}

\newcommand{\rockstar}{\textsc{Rockstar}}
\newcommand{\ctrees}{\textsc{Consistent Trees}}
\newcommand{\vmax}{v_\mathrm{max}}
\newcommand{\bpl}{BolshoiP}
\newcommand{\smdpl}{SMDPL}
\newcommand{\mdpl}{MDPL}

\usepackage{color} 
\definecolor{purple}{RGB}{160,32,240}
\usepackage {comment}
\usepackage{url}

\def\msun{\mbox{M$_{\odot}$}}

\def\lb{\mbox{$\lambda_{\rm B}$}}
\def\lp{\mbox{$\lambda_{\rm P}$}}

\def\spinB{\mbox{$\lb= J(\sqrt{2}M_{\rm vir}R_{\rm vir}V_{\rm vir})^{-1}$}}

\def\spinP{\mbox{$\lp= J|E|^{-1/2}G^{-1}M^{-5/2}$}}

\def\nh{\mbox{$n_{\rm h}$}}
\def\nsub{\mbox{$n_{\rm sub}$}}

\def\mvir{\mbox{$M_{\rm vir}$}}
\def\mvirtw{\mbox{$M_{\rm vir,12}$}}
\def\cvir{\mbox{$c_{\rm vir}$}}

\def\mpeak{\mbox{$M_{\rm peak}$}}

\def\rhom{\mbox{$\rho_{m}$}}

\def\vmax{\mbox{$V_{\rm max}$}}
\def\vpeak{\mbox{$V_{\rm peak}$}}
\def\msub{\mbox{$M_{\rm sub}$}}
\def\macc{\mbox{$M_{\rm acc}$}}
\def\Nsub{\mbox{$\langle N_{\rm sub}(>\msub|\mvir)\rangle$}}
\def\Phisub{\mbox{$\Phi_{\rm sub}(\msub|\mvir)$}}

\def\vsub{\mbox{$V_{\rm sub}$}}
\def\vacc{\mbox{$V_{\rm acc}$}}
\def\Nvsub{\mbox{$\langle N_{\rm sub}(>\vsub|\vmax)\rangle$}}
\def\Vhisub{\mbox{$\Phi_{\rm sub}(\vsub|\vmax)$}}

\def\mdyn{\mbox{$M_{\rm vir,dyn}$}}

\def\zsimtoone{\mbox{$z \sim 1$}}

\def\zgrtone{\mbox{$z\grtsim 1$}}

\defcitealias{RDA12}{RDA12}
\def\ltsima{$\; \buildrel < \over \sim \;$}    
\def\lesssim{\lower.5ex\hbox{\ltsima}}           
\def\gtsima{$\; \buildrel > \over \sim \;$}    
\def\grtsim{\lower.5ex\hbox{\gtsima}}           

\title[Halo Demographics with Planck Parameters]
{Halo and Subhalo Demographics with Planck Cosmological Parameters: Bolshoi-Planck and MultiDark-Planck Simulations}
\author[]{Aldo Rodr\'iguez-Puebla$^1$\thanks{rodriguez.puebla@gmail.com}, Peter Behroozi$^{2,3}$, Joel Primack$^4$, Anatoly Klypin$^5$, \newauthor Christoph Lee$^4$, Doug Hellinger$^4$ \\
$^1$Department of Astronomy and Astrophysics, University of California, Santa Cruz, CA 95064, USA \\
$^2$Astronomy and Physics Departments and Theoretical Astrophysics Center, University of California, Berkeley, Berkeley CA 94720, USA \\
$^3$Hubble Fellow \\
$^4$Physics Department, University of California, Santa Cruz, CA 95064, USA \\
$^5$Department of Astronomy, New Mexico State University, Las Cruces, NM 88001, USA \\
}

\date{Released 2016 Xxxxx XX}

\pagerange{\pageref{firstpage}--\pageref{lastpage}} \pubyear{2016}

\begin{document}

\label{firstpage}

\maketitle

\begin{abstract}
We report and provide fitting functions for the abundance of dark matter halos and subhalos as a function of mass, circular velocity, and redshift from the new Bolshoi-Planck and MultiDark-Planck $\Lambda$CDM cosmological simulations, based on the Planck parameters. We also report halo mass accretion rates and concentrations. We show that the higher cosmological matter density of the Planck parameters compared with the WMAP parameters leads to higher abundance of massive halos at high redshifts. We find that the median halo spin parameter \spinB\ is nearly independent of redshift, leading to predicted evolution of galaxy sizes that is consistent with observations, while the significant decrease with redshift in median \spinP\ predicts more decrease in galaxy sizes than is observed.  Using the Tully-Fisher and Faber-Jackson relations between galaxy velocity and mass, we show that a simple model of how galaxy velocity is related to halo maximum circular velocity leads to increasing overprediction of cosmic stellar mass density as redshift increases beyond \zsimtoone, implying that such velocity-mass relations must change at \zgrtone.  By making a realistic model of how observed galaxy velocities are related to halo circular velocity, we show that recent optical and radio observations of the abundance of galaxies are in good agreement with our $\Lambda$CDM simulations.  Our halo demographics are based on updated versions of the \rockstar\ and \ctrees\ codes, and this paper includes appendices explaining all of their outputs.  This paper is an introduction to a series of related papers presenting other analyses of the Bolshoi-Planck and MultiDark-Planck simulations.
\end{abstract}


\section{Introduction}

\begin{figure}
	\vspace*{-85pt}
	\hspace*{-5pt}
         \includegraphics[height=3.5in,width=3.5in]{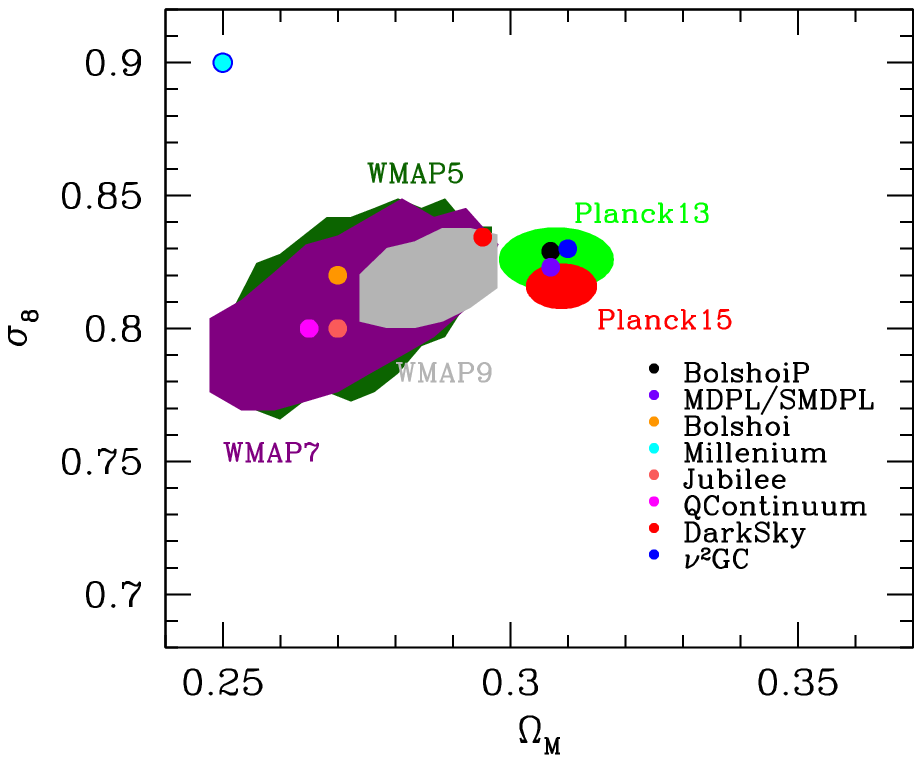}   
		\caption{Observational constraints on $\sigma_8$ and $\Omega_{\rm M}$ 
		compared to values assumed in cosmological $N-$body simulations.  The observations plotted are as follows: WMAP5+BAO+SN \citep{Hinshaw+09}, WMAP7+BAO+$H_0$ \citep{Jarosik+11}, WMAP9+eCMB+BAO+$H_0$ \citep{Hinshaw+13}, Planck13+WP+highL+BAO \citep{PlanckC13}, and Planck15+TT,TE,EE+lowP+lensing+ext \citep{PlanckC15}. 
 	}
	\label{s8-OmM.pdf}
\end{figure}

In the $\Lambda$CDM standard modern theory of structure formation in the universe, galaxies populate dark matter halos and subhalos.  The demographics of these halos as a function of redshift are thus an important input to the prediction of the properties and distribution of galaxies.  A number of large cosmological simulations have now been run \citep[see e.g.][]{Kuhlen12}, although many cover large volumes but with resolution too low to identify all dark matter halos that host most galaxies.  The mass resolution required to do this is $\lesssim 10^8 h^{-1} M_\odot$, and the force resolution should be $\lesssim 1 h^{-1}$ kpc.
High-resolution cosmological dark matter simulations that are particularly useful for studying galaxy hosts include the Millennium simulations \citep{Millennium,Millennium2,MillenniumXXL}, Bolshoi \citep{Klypin+2011}, MultiDark \citep{Prada+2012,Riebe}, Jubilee \citep{Jubilee}, DarkSky \citep{DarkSky}, Q Continuum \citep{Heitmann15}, $\nu^2$GC \citep{v2GC}, and  Bolshoi-Planck and MultiDark-Planck \citep{Klypin+2014} simulations.  
Figure 1 shows the WMAP5/7/9 \citep{WMAP9C} and Planck 2013 \citep{PlanckC13} and  Planck 2015 \citep{PlanckC15}  cosmological parameters $\sigma_8$ and $\Omega_M$, and the cosmological parameters adopted for these simulations.  The Millennium simulations used the first-year (WMAP1) parameters \citep{WMAP1}; the Bolshoi, Q Continuum, and Jubilee simulations used the WMAP5/7 cosmological parameters; while the $\nu^2$GC and Bolshoi-Planck simulations used the Planck 2013 parameters, and the DarkSky simulations used parameters between WMAP9 and Planck 2013.

In this paper we use the  \rockstar\  halo finder \citep{ROCKSTAR} and \ctrees\ \citep{CTrees} to analyze results for the recent Bolshoi-Planck (\bpl),
Small MultiDark-Planck (\smdpl) and MultiDark-Planck (\mdpl) simulations based on the 2013 Planck cosmological parameters \citep{PlanckC13} and compatible with the Planck 2015 parameters \citep{PlanckC15}.  
The \bpl, \smdpl\ and \mdpl\ simulations are not the largest of the new high-resolution simulations, but they do have the advantage that they have been analyzed in great detail, and all of these analyses are being made publicly available.  In addition, in this paper we show the effects of the change from the WMAP5/7 to the Planck 2013 cosmological parameters.

In this paper we focus on the scaling relations of several basic halo properties, updating their scaling relations as a 
 function of redshift for the Planck cosmological parameters as well as the redshift evolution of halo/subhalo number densities. For the majority  of these halo properties we report fitting functions that can be very useful not only to gain insight about the halo/subhalo population but also for the galaxy-halo connection and thus for galaxy evolution. Indeed, techniques such as subhalo abundance matching and  halo occupation distribution models require as inputs the halo/subhalo number densities. Furthermore, simplified prescriptions for the evolution of dark matter halo properties are ideal tools for people interested in understanding average properties of halos and the galaxies that they host.   

Here we analyze \textit{all} dark matter halos and subhalos found by \rockstar, and do not just focus on those that satisfy some criteria for being ``relaxed" or otherwise ``good," in contrast to some earlier studies of dark matter halo properties \citep[e.g.,][]{Bett+2007,Maccio+2007,Ludlow+2014}.   The reason is that all sufficiently massive halos are expected to host galaxies or, for the more massive ones, groups or clusters of galaxies.  

This paper is an introduction to a series of papers presenting additional analyses of the Bolshoi-Planck and MultiDark-Planck simulations.  The statistics and physical meaning of halo concentration are discussed in detail in \citet{Klypin+2014}, which is also an overview of the Bolshoi-Planck and MultiDark-Planck simulations, including BigMultiDark simulations in $(2.5 h^{-1} {\rm Gpc})^3$ volumes that we do not discuss here since they are mainly useful for statistics of galaxy clusters.   The Stellar Halo Accretion Rate Co-evolution (SHARC) assumption---i.e., that the star formation rate of central galaxies on the main sequence of star formation is proportional to their host halo's mass accretion rate---was explored in \citet{RP16}, which used abundance matching based on the Bolshoi-Planck simulation.  That paper showed that SHARC is remarkably consistent with the observed galaxy star formation rate out to $z\sim4$ and that
the $\sim0.3$ dex dispersion in the halo mass accretion rate is consistent with the observed small dispersion of the star formation rate about the main sequence. 
The clustering properties of halos and subhalos is the subject of Rodriguez-Puebla et al. 2016b (in prep.).
How properties of dark matter halos vary with the density of their environment on length scales from 0.5 to 16 $h^{-1}$ Mpc is discussed in Lee et al. (2016a, in prep.), which shows among other things that halos in low-density regions experience lower tidal forces and have lower spin parameters, and that a large fraction of lower-mass halos in high-density regions are ``stripped," i.e. their mass at $z=0$ is less than that of their progenitors at higher redshifts.  Another paper (Lee et al., 2016b, in prep.) studies the causes of halo stripping and properties of such stripped halos.  Further papers comparing with observations are also in preparation, along with mock galaxy catalogs based on Bolshoi-Planck.

This paper is organized as follows: \S2 discusses the simulations and how we define the halo mass.  \S3 describes the key scaling relations for distinct halos (i.e., those that are not subhalos) and gives figures and fitting formulas for maximum halo circular velocity (\S3.1), halo mass accretion rates (\S3.2) and mass growth (\S3.3).  \S4 discusses halo (\S4.1) and subhalo (\S4.2) number densities, and the number of subhalos as a function of their host halo mass (\S4.3).  \S5 presents the halo and subhalo velocity functions.  \S4 and \S5 also compare the Planck cosmology halo mass and velocity functions with those from the WMAP5/7 cosmological parameters. \S6 discusses the dependence of halo concentration and spin on mass and redshift.  \S7 discusses the evolution of the Tully-Fisher and Faber-Jackson relations between halo circular velocity $V_{\rm max}$ and the stellar mass of the central galaxies in these halos. \S8 compares the halo velocity function with the galaxy velocity function from optical and radio observations. 
\S9 summarizes and briefly discusses the key results in this paper.  Appendix A is an overview of the \ctrees\ merger tree information and halo catalogs and Appendix B summarizes the \rockstar\ and \ctrees\ fields.

\section{The simulations}

\makeatletter{}\begin{table*}
 \begin{minipage}{16.cm}
\caption{Numerical and cosmological parameters for the simulations analyzed in this paper.
  The columns give the simulation identifier on the CosmoSim website, 
  the size of the simulated box in $h^{-1}\,{\rm Gpc}$,
  the number of particles, 
  the mass per simulation particle $m_p$ in units $h^{-1}\,M_\odot$,
  the Plummer equivalent gravitational softening length $\epsilon$  in units of
physical  $h^{-1}\,{\rm kpc}$,
   the adopted values for
  $\Omega_{\rm{Matter}},\, \Omega_{\rm{Baryon}},\, \Omega_{\Lambda}$,
  $ \;\sigma_8$, the spectral
index $n_s$, and the Hubble constant $H_0$ in ${\rm km/s/Mpc}$.
The references for these simulations are (a) \citet{Klypin+2014},  (b) \citet{Klypin+2011}, (c) \citet{Prada+2012}.}
\begin{tabular}{ l | c | c |  c | c | c | c | c | c | c | c | l | l |l }
\hline  
Simulation & box &  particles  & $m_p$ & $\epsilon$ &
$\Omega_M$ & $\Omega_B$ & $\Omega_{\Lambda}$ & $\sigma_8$ & $n_s$ & $H_0$ & Code & Ref.

\tabularnewline
  \hline   
BolshoiP   & $0.25$   &  $2048^3$   &  $1.5 \times 10^{8}$  &  $1.0$ &
$0.307$  & $0.048$   &  $0.693$  &  $0.823$  &  $0.96$  &  $67.8$ & {\small ART} & a
\tabularnewline

SMDPL   & $0.4$   &  $3840^3$   &  $9.6 \times 10^{7}$  &  $1.5$ &
$0.307$  & $0.048$   &  $0.693$  &  $0.829$  &  $0.96$  &  $67.8$ & {\small GADGET-2} &	a
\tabularnewline

MDPL    & $1.0$   &  $3840^3$   &  $1.5 \times 10^{9}$  &  $5$ &
$0.307$  & $0.048$   &  $0.693$  &  $0.829$  &  $0.96$  &  $67.8$ & {\small GADGET-2} &	a
\tabularnewline

\hline
Bolshoi   & $0.25$   &  $2048^3$   &  $1.3 \times 10^{8}$  &  $1.0$ &
$0.270$  & $0.047$   &  $0.730$  &  $0.820$  &  $0.95$  &  $70.0$ & {\small ART} & b
\tabularnewline

MultiDark   & $1.0$   &  $2048^3$   &  $8.7 \times 10^{9}$  &  $7.0$ &
$0.270$  & $0.047$   &  $0.730$  &  $0.820$  &  $0.95$  &  $70.0$ & {\small ART} & c
\tabularnewline

\hline
\end{tabular}
\label{simtable}

\end{minipage}

\end{table*}

The cosmological parameter values for the Bolshoi-Planck and MultiDark-Planck simulations are
$\Omega_{\Lambda,0}=0.693,\Omega_{\rm M,0}=0.307,\Omega_{\rm B,0}=0.048, h=0.678, n_s=0.96$  and $\sigma_8=0.823$.  The parameters are the same for the MultiDark-Planck simulations except for $\sigma_8=0.829$.  Simulation volumes, resolutions and other parameters of the Bolshoi and MultiDark simulations with WMAP7/9 parameters, and the new Bolshoi-Planck and the $(400 h^{-1} {\rm Mpc})^3$  and $(1 h^{-1} {\rm Gpc})^3$  MultiDark-Planck simulations are summarized in Table 1.  The details about the number and redshift distribution of the saved timesteps of these simulations are given in Appendix A.
Outputs from these simulations are available online at the CosmoSim website.\footnote{\url{https://www.cosmosim.org/cms/simulations/multidark-project/} with explanations in \citet{Riebe} and at \url{https://www.cosmosim.org/cms/documentation/introduction/whats-different/} for users of the earlier site \url{http://www.multidark.org/}.}  Entire \rockstar\ and  \ctrees\ outputs are downloadable from another website.\footnote{\url{http://hipacc.ucsc.edu/Bolshoi/MergerTrees.html}}

\begin{figure}
	\vspace*{-130pt}
	\includegraphics[height=4.5in,width=4.5in]{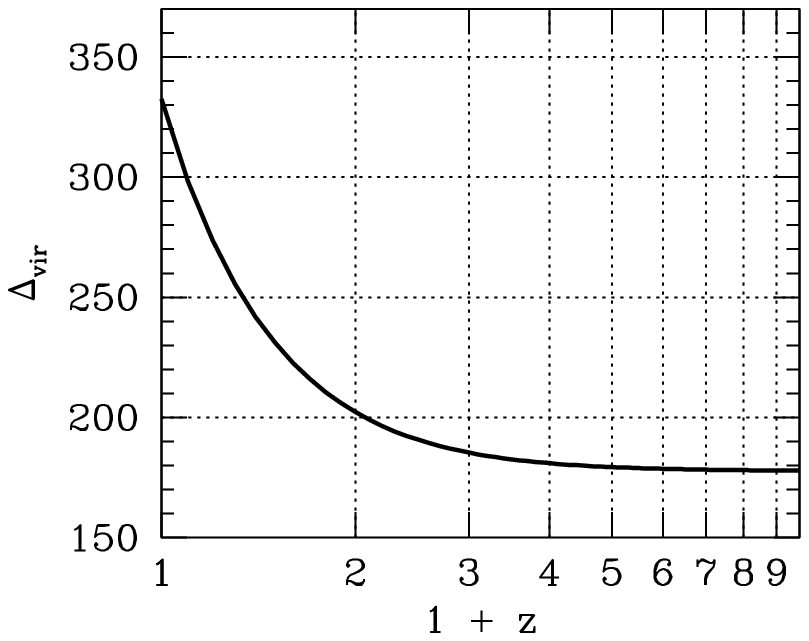}
	\vspace*{-30pt}
		\caption{Virial overdensity $\Delta_{\rm vir}$ 
		given by the spherical collapse model \citep{BryanNorman}.
		The value of the virial overdensity at $z=0$  is   $\Delta_{\rm vir} = 333$ for 
		the Bolshoi-Planck cosmological parameters, while 
		for large $z$ it approaches to $\Delta_{\rm vir} = 178$. 
 	}
	\label{f1a}
\end{figure}

In this paper we define halo masses by using  
spherical overdensities according to the redshift-dependent virial overdensity $\Delta_{\rm vir}(z)$ 
given by the spherical collapse model, for which \citet{BryanNorman} give the fitting formula
$\Delta_{\rm vir}(z) = (18\pi^2 + 82x - 39x^2)/\Omega(z)$, where $\Omega(z) $ is the ratio of mean matter density $\rho_{\rm m}$ to critical density $\rho_{\rm c}$ at redshift $z$, 
and $x \equiv \Omega(z) -1$. 
Figure \ref{f1a} shows the redshift dependence of $\Delta_{\rm vir}$ for the cosmology of the
Bolshoi-Plank simulation. The value of the virial overdensity at $z = 0$ is $\Delta_{\rm vir} = 333$,
while for large $z$ it asymptotes to $18\pi^2 =178$.  The virial radius $R_{\rm vir}$ of a halo of virial mass $\mvir$ is defined as the radius within which the mean density is $\Delta_{\rm vir}$ times the mean matter density $\rho_{\rm m} = \Omega_M \rho_{\rm c}$ at that redshift.
Then the virial halo mass is 
\begin{equation}
\mvir = \frac{4\pi}{3} \Delta_{\rm vir} \rho_{\rm m} R_{\rm vir}^3 .
\label{Mvir}
\end{equation}
Another common choice employed to define halos is the radius $R_{\rm 200c}$ enclosing 200 times critical density, with corresponding halo mass
\begin{equation}
M_{\rm 200c} = \frac{4\pi}{3} 200 \rho_{\rm c} R_{\rm 200c}^3 ,
\end{equation}
which was used in \citet{Klypin+2014}.  Yet another common choice is $R_{\rm 200m}$ enclosing 200 times mean density, with corresponding halo mass $M_{\rm 200m}$.  Although we use $M_{\rm vir}$ in this paper, the \rockstar/\ctrees\ analyses of the Bolshoi-Planck and MultiDark-Planck simulations described in the Appendices include outputs for both $M_{\rm vir}$ and $M_{\rm 200m}$.

We note that \citet[][ see also \citealp{Zemp+2014}]{Diemer+2013} argued that much of the mass evolution of dark matter halos is an artifact caused by the changing radius of the halo as the mean cosmic matter density  $\rho_{\rm m}$ declines as the universe expands.  They call this phenomenon ``pseudoevolution,"  since the dark matter distribution in the interior of most halos hardly changes at low redshift \citep{Prada+2006,Diemand+2007,Cuesta+2008}.  Recently \citet{More+2015} proposed that the best physically-based definition of halo radius is the ``splashback radius'' $R_{\rm sp} \approx 2 R_{\rm 200m}$, where there is typically a sharp drop in the density.  Using this definition, there is actually more halo mass increase than for $R_{\rm vir}$ or $R_{\rm 200m}$.  This is discussed further in \citet{RP16}, where we argue that for purposes like predicting galaxy star formation, the $R_{\rm vir}$ definition used here works fine.

\section{Basic scaling relations for distinct halos}

\begin{figure*}
	\vspace*{-250pt}
	\hspace*{-20pt}
	\includegraphics[height=6.5in,width=6.5in]{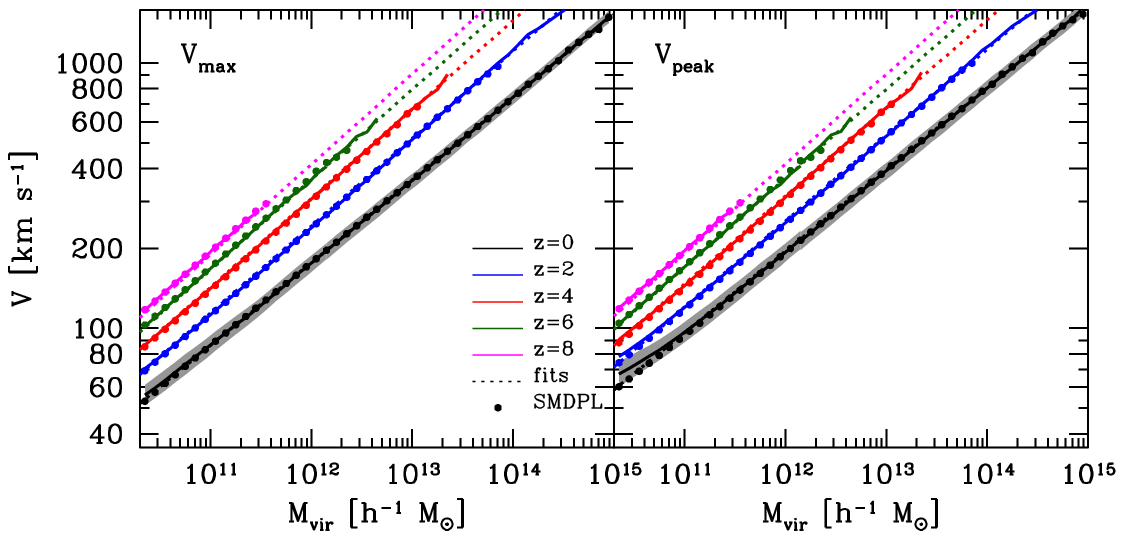}
		\caption{{\bf Left Panel:}
		Maximum halo circular velocity, \vmax, as a function of \mvir\ at $z = 0, 2, 4, 6$ and $8$. 
		Medians are shown as the solid lines for the \bpl\ and \mdpl\
		simulations, filled circles are the medians of the \smdpl\ simulation. 
		At $z=0$ the grey band is the $68\%$
		range of the maximum  circular velocity. The dotted lines show the fits to the simulation. 
		A single power law is able to reproduce the results from the simulation.
		The slopes are approximately independent of redshift with a value of $\sim1/3$. 
		{\bf Right Panel:} The highest maximum circular velocity reached  
		along the main progenitor branch, \vpeak, as a function of \mvir\ at $z = 0, 2, 4, 6$ and $8$.
		Similarly to the \vmax\ panel, medians are shown as the solid lines. At $z=0$ the grey band is the 
		$1\sigma$ $(68\%)$ range of the maximum  circular velocity. The dotted lines show the 
		fits to the simulation. Also, the slopes are approximately independent of redshift with a 
		value of $\sim1/3$. 
 	}
	\label{f1}
\end{figure*}

Galaxies form and evolve in dark matter halos, and it is expected that visible galaxies are hosted by all halos in the mass range 
$\mvir\grtsim10^{10.2}h^{-1}\msun$ where
halos can be resolved with at least  $\grtsim 100$ particles in the \bpl\ and \smdpl\ simulations. Therefore,
the statistical properties of dark matter halos, which can be studied in great detail in high resolution numerical $N-$body 
simulations, can provide hints on the nature of galaxy properties and spatial distributions. 
In this Section, we report dark matter halo velocity and mass and their scaling relations.  Dependence on mass and redshift of halo concentration and spin are discussed in 
\S\ref{C-lambda}.

\subsection{Maximum halo circular velocity}
\label{vmax_section}

As usual, the circular velocity is defined as $V_{\rm circ} \equiv \sqrt{G M(<r)/r}$, where $M(<r)$ is the halo mass enclosed by a sphere of radius $r$.  Dark matter halos have circular velocity that grows from 0 at $r=0$ to a maximum value \vmax\ at a radius $R_{\rm max}$ that is usually considerably less than $R_{\rm vir}$.  (\S\ref{concentrations} shows that for the NFW radial halo mass distribution, $R_{\rm max} = 2.1626\times\ R_s$.)  Because \vmax\ characterizes the inner halo, it may correlate better with the properties of the central galaxy than \mvir\ does.
The left panel of Figure \ref{f1} shows the medians of the
maximum halo circular velocity, \vmax, as a function of \mvir\ at $z = 0, 2, 4, 6$ and $8$, solid lines.
The grey band at $z=0$ shows the $68\%$ range of the maximum  circular velocity, i.e., 
the halo distribution between the 16th and 84th percentiles. We find that the $68\%$ range 
of the distribution is approximately
independent of redshift and halo mass, with a value of $\sim 0.05$ dex. In general, the \vmax--\mvir\ relation
follows a power law-fit at all redshifts and over the mass range where 
 we can resolve distinct halos in the Bolshoi-Planck simulations, $\mvir\sim10^{10.2}\msun$. 
 To a good
approximation, the \vmax--\mvir\ slope is given by  $\alpha\sim 1/3$, as expected from  spherical collapse.
In reality, however, the slope depends slightly on redshift as we will quantify below. 

Distinct halos can lose mass due to stripping events as a result of interactions with other halos. 
In consequence, the maximum halo circular velocity \vmax\ can significantly decrease (this will be discussed
in more detail in Lee et al 2016b). 
This reduction in \vmax\ can introduce an extra source of uncertainty when relating galaxies to dark matter halos, since it is expected that 
stripping would affect halos more significantly than the central galaxies deep inside them. Therefore, in the case of stripped halos, the correlation
between the present \vmax\ of the halo and galaxy stellar mass/luminosity is not trivial. Indeed, 
\citet{Moster+2010} and \citet{Reddick+2013} found that the highest maximum circular velocity reached  
along the halo's main progenitor branch, \vpeak, is a better halo proxy for galaxy stellar 
mass/luminosity. 
For these reasons 
we find it useful to report the \vpeak--\mvir\ relation in this paper. 

The right panel of
Figure \ref{f1} shows the redshift evolution of the highest maximum circular velocity reached  
along the main progenitor branch, \vpeak, as a function of \mvir\ at $z = 0, 2, 4, 6$ and $8$. As for  \vmax--\mvir, medians are shown as the solid lines. The grey band at $z=0$ is the $68\%$ of the 
distribution. We find that the slope depends more on redshift than for the  \vmax--\mvir\ relation. 
The differences in the slopes, especially at lower redshifts, are a consequence of tidal stripping events, as mentioned above. 
On the other hand, we find that the scatter is independent of redshift and mass and of the order of $\sim0.05$ dex, similarly to \vmax--\mvir.

\begin{figure*}
	\vspace*{-80pt}
	\includegraphics[height=5.5in,width=5.5in]{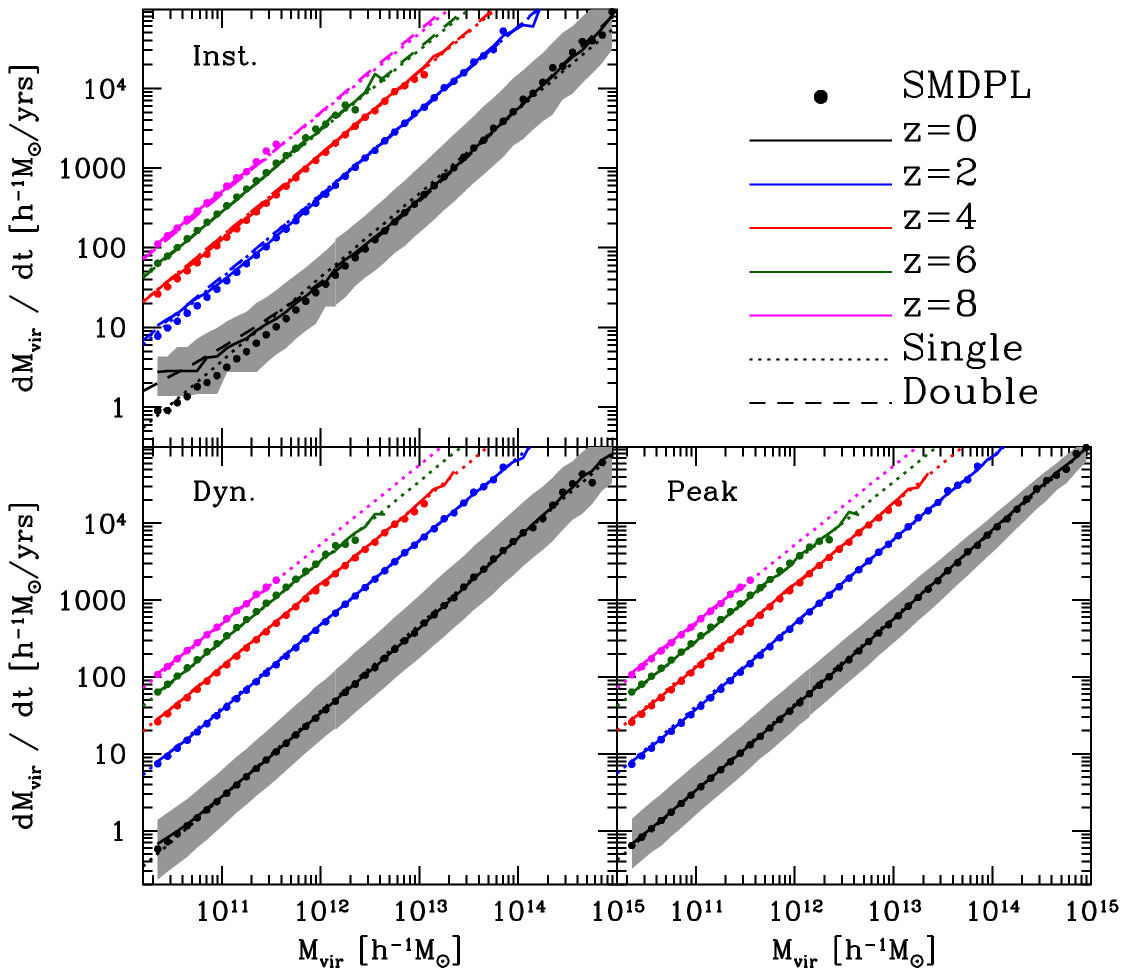}
		\caption{Halo mass accretion rates as a function of \mvir\ at $z = 0, 2, 4, 6$ and $8$. 
		Medians are shown as the solid lines for the \bpl\ and \mdpl\
		simulations, and filled circles are the same for the \smdpl\ simulation. At $z=0$ the grey band is the $68\%$
		range of halo mass accretion rates for the \bpl\ and \mdpl\
		simulations. The dotted lines show the fits to the simulations
		when using a power law, Equation (\ref{dmhdt-mh}). 
		{\bf Upper Panel:} Instantaneous halo mass accretion rates. The long dashed line shows the fits  
		using a double power law, Equation (\ref{dmhdt-mh-dbl}), for the \bpl\ and \mdpl\ simulations.  
		{\bf Bottom Left Panel:} Halo mass accretion rates averaged over a dynamical time.
		A single power law reproduces the results from the simulations. 
		The slopes are approximately independent of redshift with a value of $\sim1.1$. 
		{\bf Bottom Right Panel:} Mass accretion rates of \mpeak\ averaged 
		from the current halo's redshift, $z$, to $z+0.5$. 
		As for the dynamical time averaged accretion rates,
		a single power law again reproduces the results from the simulations. 
		The slopes are approximately independent of redshift with a value of $\sim1.1$. 
 	}
	\label{f2}
\end{figure*}

The dotted lines in Figure \ref{f1} show power-law fits to the \bpl, \smdpl\ and \mdpl\
simulations both for the
\vmax--\mvir\ (left panel) and \vpeak--\mvir\ (right panel) relationships. We motivate 
the power law-fits based on the well known results for isothermal dark matter halo profiles. 
For a singular isothermal sphere the circular velocity 
(assumed to be independent of radius inside the halo) evolves with redshift as
\begin{equation}
\vmax\ \propto\ \left[ \mvir E(z)\right]^{1/3},
\label{iso_shpere}
\end{equation}
where $E(z)$ is the expansion rate $H/H_0$ for a flat universe:
\begin{equation}
E(z) = \sqrt{ \Omega_{\Lambda,0}+\Omega_{\rm m,0}(1+z)^3}.
\label{E(z)}
\end{equation}
Of course this is a simplification since the mass profile of dark matter halos 
is markedly different from an isothermal sphere as previous studies based on
high-resolution $N$-body simulations have shown 
\citep[for a recent discussion see:][]{Klypin+2015}. However, the general structure of  
Equation (\ref{iso_shpere}) can be useful to suggest fitting functions for redshift evolution. 
Based on this, we assume the following parametric form for the redshift evolution  both for the
\vmax--\mvir\ and \vpeak--\mvir\ relationships
	\begin{equation}
			V(\mvir,z) = \beta(z) \left[ \mvirtw  E(z) \right] ^ { \alpha(z) },
		\label{vmax-mh}
	\end{equation}
where $\mvirtw \equiv \mvir/(10^{12}h^{-1}\msun)$.
Our fitting functions for \vmax--\mvir, with $a$ representing the scale factor $a = 1/ (1 + z)$, are
\begin{equation}
		\alpha(z) =  0.346 - 0.059 a +  0.025 a^2,
	\end{equation}
and
	\begin{equation}
		\log \beta(z) =  2.209 +  0.060 a -  0.021 a^2,
	\end{equation}
while the fitting functions for \vpeak--\mvir\ are
\begin{equation}
		\alpha(z) =  0.346 -  0.080 a +  0.042 a^2,
	\end{equation}
and
	\begin{equation}
		\log \beta(z) =  2.205 +  0.150 a - 0.063 a^2 .
	\end{equation}

\subsection{Halo mass accretion rates}

Halo mass accretion is responsible for controlling the rate at which  
the baryonic mass, $M_{b}$, is deposited in galaxies.
In the past, instantaneous halo mass accretion rates, $d\mvir / dt$, have been studied in great detail based on
the Extended Press-Schechter formalism \citep{Kauffmann+1993,Lacey+1993,Somerville+1999,vandenBosch2002,Neistein+2006,Zhang+2008,Firmani+2013} 
or high-resolution $N-$body simulations \citep{Wechsler+2002,Cohn+2005,McBride+2009,Tweed+2009,Fakhouri+2010,Srisawat+2013} or either both
\citep{Parkinson+2008,Jiang+2014,vandenBosch+2014}. In addition to
instantaneous halo mass accretion rates $d\mvir / dt$ calculated between stored simulation timesteps, in this paper we extend the above
work by also studying halo mass accretion rates averaged over the dynamical time $d\mdyn / dt$, defined as
	\begin{equation}
		\big\langle\frac{d\mvir}{dt}\big\rangle_{\rm dyn} = \frac{\mvir(t) - \mvir(t-t_{\rm dyn})}{t_{\rm dyn}},
	\label{tdyn}
	\end{equation}
where the dynamical time of the halo is $t_{\rm dyn}(z) = [G \Delta_{\rm vir}(z) \rho_{\rm crit}(z)]^{-1/2}$.  The ratio of the dynamical time to the  Hubble time $t_H = H^{-1}$ is  
$t_{\rm dyn}/t_H = [(8\pi)/(3 \Delta_{\rm vir})]^{1/2}$, which equals 0.16 at $a=0$ and asymptotes to 0.22 at high redshift.  We also report halo mass accretion rates of the 
maximum mass  \mpeak\ reached  along the main progenitor branch 
averaged from the current halo's redshift, $z$, to $z+0.5$, $d\mpeak / dt$. For more details the reader 
is referred to Appendix B4.

Figure \ref{f2} shows the medians of halo mass accretion rates at $z = 0, 2, 4, 6$ and $8$. 
The upper panel shows the instantaneous halo mass accretion rate, $d\mvir / dt$, while 
the left and right bottom panels show halo mass accretion rates averaged over a dynamical time, $d\mdyn / dt$,
and \mpeak\ halo accretion rates, $d\mpeak / dt$, respectively. The grey band at $z=0$ in all the panels shows the distribution of
halo mass accretion rates between the 16th and 84th percentiles, i.e., the $68\%$ of the distribution. We find that the dispersion is roughly
independent of redshift and in slightly dependent on halo mass. 
 The dispersion for $d\mvir / dt$ is roughly $\sim0.35$ dex, while 
 for $d\mdyn / dt$ is $\sim0.3$ dex, and for $d\mpeak / dt$ is $\sim0.25$ dex. The $d\mvir / dt - \mvir$ relations
follow a power law-fit, especially at high redshifts. As a crude estimation, the slope 
for the $d\mvir / dt - \mvir$ relations is  $\alpha\sim 1.1$, consistent with previous studies based on 
the Press-Schechter formalism \cite[e.g., ][]{Neistein+2006} and the Millennium and Millennium-II high resolution $N-$body simulations
 \citep{Fakhouri+2010}.

\begin{table*}
	\caption{Best fit parameters for the $d\mvir/dt-\mvir$ relationships.}
	\begin{center}
		\begin{tabular}{c c c c c  c c }
			\hline
			\hline	
			$d\mvir/dt$ [$h^{-1}$\msun/yrs] & $\alpha_0$ & $\alpha_1$ &  $\alpha_2$ & $\beta_0$ & $\beta_1$  &  $\beta_2$  \\
			\hline
			\hline
			 Instantaneous & 0.975 & 0.300 & -0.224 & 2.677 & -1.708 & 0.661 \\
			\hline
			Dynamical averaged  & 1.000 & 0.329 & -0.206 & 2.730 & -1.828 & 0.654 \\
			\hline
			Peak & 0.997 & 0.328 & -0.200 & 2.711 & -1.739 & 0.672 \\
			\hline
			\hline
		\end{tabular}
		\end{center}
	\label{T2}
\end{table*}

The dotted lines in Figure \ref{f4} show power-law fits to the simulations for the halo mass accretion rates, given by
	\begin{equation}
			\frac{d\mvir}{dt} = \beta(z)  \mvirtw^ { \alpha(z) } E(z),
		\label{dmhdt-mh}
	\end{equation}
where 
	\begin{equation}
		\log \beta(z) =  \beta_0 +  \beta_1 a +  \beta_2 a^2,
	\end{equation}
and 
	\begin{equation}
		\alpha(z) =  \alpha_0 +  \alpha_1 a +  \alpha_2 a^2.
	\end{equation}
Table \ref{T2} lists the best fit parameters for the $d\mvir / dt - \mvir$ relations. 
Power-law fits can provide an accurate description for both $d\mdyn / dt$ 
and $d\mpeak / dt$ for the three simulations. 

As can be observed in the upper panel of Figure \ref{f2}, however, a power-law fit is a poor description of the instantaneous halo mass accretion rates, especially for the \bpl and \mdpl\ simulations at low masses and low redshifts.  
In order to find a better description of the instantaneous halo mass accretion rates for  
the \bpl\ and \mdpl\  simulations we use a double power-law fit  
	\begin{equation}
			\frac{d\mvir}{dt} = \beta(z)[\mvirtw^{\alpha(z)} +   \mvirtw^{ \gamma(z) }] E(z),
		\label{dmhdt-mh-dbl}
	\end{equation}
where  the normalization is given by
	\begin{equation}
		\log \beta(z) = 2.437 - 1.857\times a +  0.685\times a^2,
	\end{equation}
and the powers $\alpha(z)$ and $\gamma(z)$ are given respectively by
	\begin{equation}
		\alpha(z) =  1.120 -  0.609\times a +  0.097\times a^2,
	\end{equation}
 and
	\begin{equation}
		\gamma(z) =  0.917 +  0.845\times a -  0.532\times a^2.
	\end{equation}
The dashed lines in Figure \ref{f2} show this fit to the simulations. 

Finally, based on the above definitions of halo accretion rates, 
the rate at which the cosmological baryonic inflow material is  accreted into the dark matter halo 
is calculated as $dM_{\rm c,b} / dt = f_{\rm c,b}\times d\mvir / dt$, where the cosmic
baryon fraction is $f_{\rm c,b} \equiv \Omega_{\rm B,0} / \Omega_{M,0} = 0.156$ for our cosmology. The rate $dM_{\rm c,b} / dt$ is an important quantity; it equals the star formation rate plus the gas outflow rate if the galaxy is in ``equilibrium'' in bathtub model terms \citep[e.g.,][and references therein]{Mitra+2015}. 
	
Galaxies can be divided into two main groups: star-forming and quiescent. Star-forming galaxies are
typically blue young disk galaxies, while many quiescent galaxies
are red old spheroids. These properties are partially determined by the mass of the dark
matter halo in which they reside but, due to complexity of the galaxy formation process, a dependence on other
halo and/or environmental properties is expected. For example, star-forming galaxies at a given redshift are known to show a tight dependence
of star formation rates on stellar mass, which is known as the ``main
sequence" of galaxy formation. The slopes and dispersions of halo mass accretion rates reported above are 
very similar to the observed dispersion and slope of the star formation rates on the main sequence. This naturally suggests that 
the halo mass accretion rate is controlling not only the baryon fraction that is entering the
galaxies, but also their star formation efficiency. The galaxy stellar-to-halo mass relation is known to be nearly independent of redshift from $z=0$ out to $z\sim 4$
\citep{Behroozi+2013L}, so the galaxy star formation rate is determined on average by
the mass accretion rate of the halo in which it resides: $d M_\ast / dt = (d M_\ast / d \mvir) (d \mvir / dt)$.  A recent paper by some of us, \citet{RP16},  made the stronger assumption that this is true halo by halo for star-forming galaxies, which we called Stellar-Halo Accretion Rate Coevolution (SHARC). 
We showed that the SHARC assumption predicts galaxy star formation rates on the main sequence that are in good agreement with observations up to $z\sim4$, and that in addition it approximately matches the small observed dispersion of $\sim0.3$ dex of the galaxy star formation rates about the main sequence.
 
 \subsection{Halo assembly}
\begin{figure}
	\vspace*{-170pt}
	\hspace*{-20pt}
	\includegraphics[height=5.2in,width=5.2in]{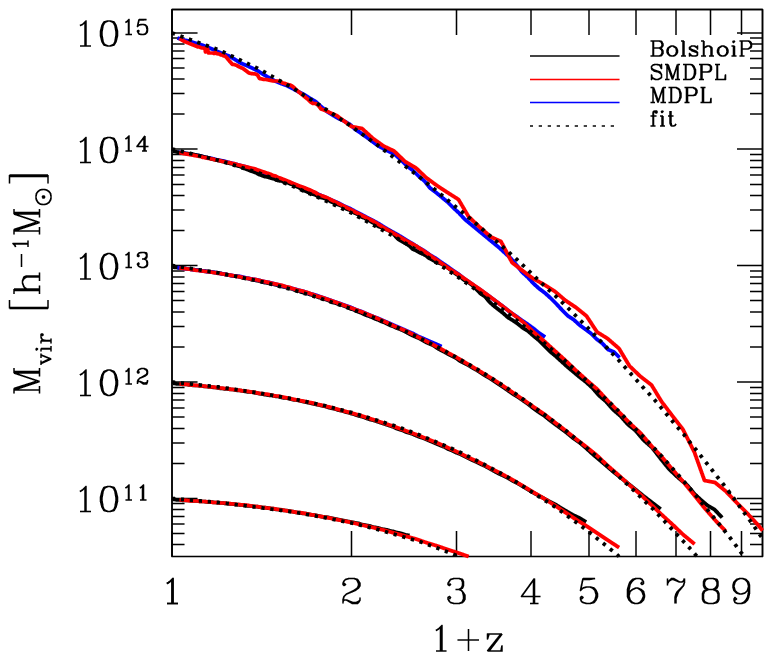}
		\caption{Median halo mass growth for progenitors $z = 0$ with
		masses of $\mvir=10^{11},10^{12},10^{13},$ and 
		$10^{14} h^{-1}$\msun, solid lines. Fits to simulations are shown
		with the dotted lines. The shaded area shows the dispersion around the medians. 
 	}
	\label{f4}
\end{figure}

\begin{figure}
	\vspace*{-170pt}
	\hspace*{-20pt}
	\includegraphics[height=5.2in,width=5.2in]{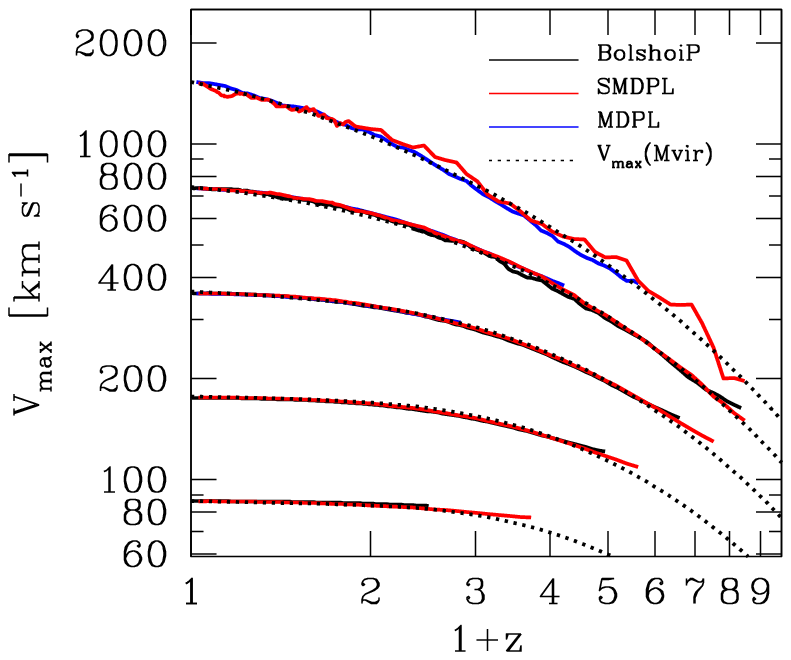}
		\caption{Median \vmax\ growth for progenitors with masses of
		$\mvir=10^{11},10^{12},10^{13},$ and 
		$10^{14}h^{-1}$\msun, solid lines. Fits to simulations are shown
		with the dotted lines.  The shaded area shows the dispersion around the medians. 
 	}
	\label{f5}
\end{figure}

Figure \ref{f4} presents the medians of the halo mass growth for progenitors at $z = 0$ with 
masses of $\mvir=10^{11},10^{12},10^{13}, 10^{14}$ and $10^{15}h^{-1}$\msun, 
for the \bpl\ (black solid line) \smdpl\ (red solid line) and \mdpl\ (blue solid line) simulations. {\color{black} In order to avoid resolution
effects and thus obtain reliable statistics we require that at every redshift at least $90\%$ of the halos can be resolved with at least  
100 particles. The first thing to note is that the three simulations agree with each other at all redshifts. From the figure it
is evident that high mass halos assembled more rapidly at later epochs than lower mass halos. 
This is consistent with the fact the slopes obtained for halo mass accretion rates are slightly greater than 1. 
For the Planck cosmology we find that $10^{12}h^{-1}$\msun\  halos formed half of their mass by  $z\sim1.2$. Progenitors
of $\mvir = 10^{13}, 10^{14}$, and $10^{15}h^{-1}$ \msun\ halos reached the mass of $10^{12}h^{-1}\msun$ at $z\sim 2.5,3.9$, and $z\sim5$
respectively. Theoretically, the characteristic mass of $10^{12}h^{-1}$\msun\ is expected to mark a transition above which 
the formation of stars in galaxies becomes increasingly inefficient. The reasons for this are that at halo masses above $10^{12}h^{-1}$\msun\ 
the efficiency at which the virial shocks can heat the gas increases \citep[e.g.,][]{Dekel+2006}, 
and the gas can be kept from cooling by energy emitted from accretion onto supermassive black holes in these high-mass halos. Thus central
galaxies in massive halos are expected to become passive systems roughly at the epoch when the halo reached the mass 
of $10^{12}h^{-1}$\msun.  Note that Figure 2 of \citet{Ludlow+2016} is similar to Figure \ref{f4}, including also a comparison to warm dark matter cosmologies.
 
In order to characterize the growth of dark matter halos we use the fitting function from \citet{Behroozi+2013}
	\begin{equation}
		\mvir (\mvir(0), z) = M_{\rm 13} (z) 10 ^{f(M_{\rm vir}(0), z)}
		\label{AMH}
	\end{equation}
where 
	\begin{equation}
		M_{\rm 13} (z) = 10^{13.6}h^{-1}\msun(1+z)^{2.755}(1+\frac{z}{2})^{-6.351}\exp{(-0.413 z)}
	\end{equation}
	
	\begin{equation}
		f(\mvir,z) = \log\left(\frac{\mvir(0)}{M_{\rm 13} (0)}\right)\frac{g(\mvir(0),1)}{g(\mvir(0,a))}
	\end{equation}

	\begin{equation}
		g(\mvir(0),a) = 1 + \exp[-3.676 (a-a_0(\mvir(0)))]
	\end{equation}
	
	\begin{equation}
		a_0(\mvir(0)) = 0.592 - \log\left[0.113\left(\frac{10^{15.7}h^{-1}\msun}{\mvir(0)}\right)+1\right].
	\end{equation}
As is emphasized in \citet{Behroozi+2013}, as opposed to other previous descriptions, the above parameterization avoids 
the problem that progenitor histories of halos with different masses cross. 

Figure \ref{f5} presents the medians of the maximum circular velocity growth for the progenitors of halos of the same masses $\mvir = 10^{11}$ to $10^{15}h^{-1}$ \msun\
described above.  
The maximum circular velocity is more directly connected to the central potential depth of the halo than the virial circular 
velocity or mass \citep[see e.g.,][]{vandenBosch+2014}, and presumably more connected to the formation of the host galaxy. 
We find that for halos of $\mvir=10^{11}h^{-1}\msun$, \vmax\ is practically constant after $z\sim2$, while for  
halos of $\mvir=10^{12}h^{-1}\msun$ and $\mvir=10^{13}h^{-1}\msun$,  \vmax\ is constant since $z\sim1$ and $z\sim0.5$, respectively. This is  
consistent with the fact that the interiors of halos hardly change at low redshifts \citep{Prada+2006,Diemand+2007,Cuesta+2008}.  
The dotted line in Figure  \ref{f5} shows the growth of \vmax\ when combining Equations (\ref{vmax-mh}) and (\ref{AMH}). 
This simple prescription reproduces accurately the redshift dependence of halo \vmax\ growth. 
}

\section{Halo and subhalo number densities}

\subsection{Distinct halo mass function}

\begin{figure*}
	\vspace*{-130pt}
	\hspace*{5pt}
	\includegraphics[height=4in,width=4in]{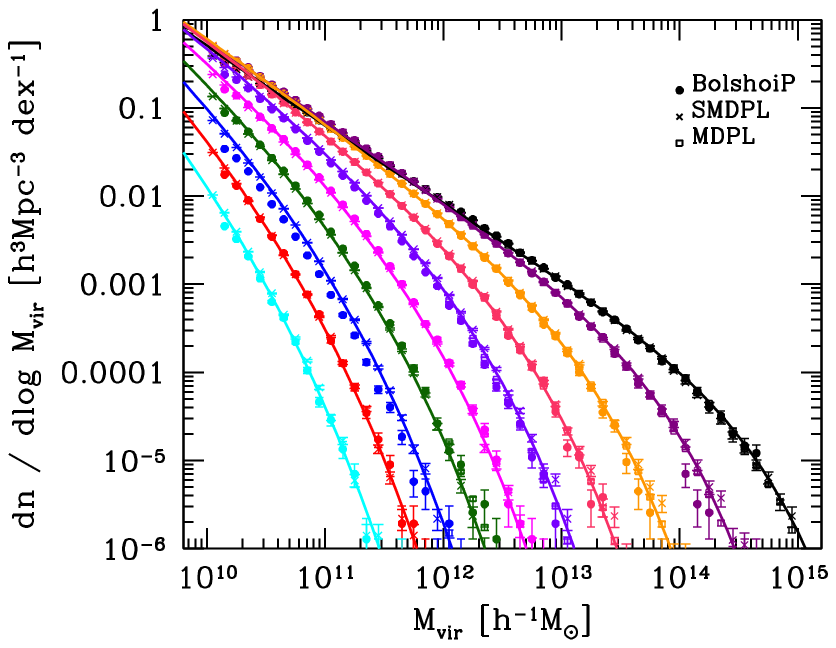}
	\hspace*{-90pt}
	\includegraphics[height=4in,width=4in]{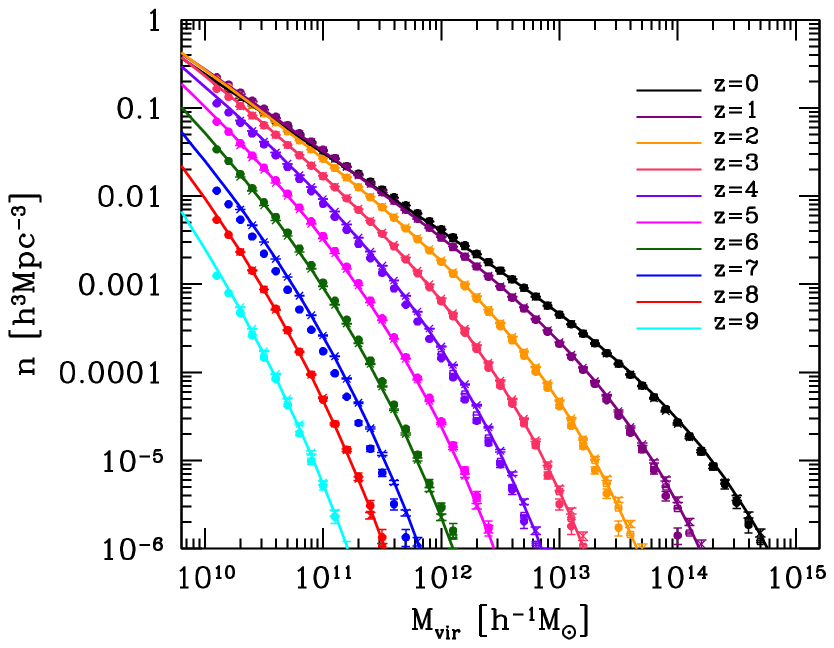}
		\caption{{\bf Left Panel:} The halo mass function from $z=0$ to $z=9$. {\bf Right Panel:}
		Cumulative halo mass function. The various solid lines show the fits to the simulations, Equation (\ref{Tinker}).
 	}
	\label{f6}
\end{figure*}

\begin{table}
	\caption{Best fit parameters to the \citet{Tinker+2008} halo mass function Equations (\ref{Tinker}) and (\ref{TinkerFit}).}
	\begin{center}
		\begin{tabular}{c c c c}
			\hline
			\hline	
			$\chi_{i}$ & $\chi_{0,i}$ & $\chi_{1,i}$ &  $\chi_{2,i}$ \\
			\hline
			\hline
			 $A$ & 0.144 & -0.011 & 0.003\\
			\hline
			$a$  & 1.351 & 0.068 & 0.006 \\
			\hline
			$b$  & 3.113 & -0.077 & -0.013 \\
			\hline
			$c$  & 1.187 & 0.009 & 0.000\\
			\hline
			\hline
		\end{tabular}
		\end{center}
	\label{T3}
\end{table}

The comoving number density of distinct halos at the mass range $\log\mvir$ and $\log\mvir+d\log\mvir$,
i.e., the halo mass function $(d\nh/d\log\mvir)$, are presented in the left panel of 
Figure \ref{f6} both for the Bolshoi-Planck and MD-Planck simulations. 
The right panel of the same figure shows the cumulative comoving number density, $\nh(>\mvir)$. 
We compare the measured halo mass function from the simulations to the analytical fitting formula reported in \citet{Tinker+2008}, which we find
provides accurate fits to the results of the Bolshoi-Planck and MD-Planck simulations at low redshifts. 
At high redshifts, however, it tends to underestimate $d\nh/d\log\mvir$  \citep[see also][]{Klypin+2014}. 

Theoretically, the comoving number density of halos at the mass range \mvir\ and $\mvir+d\mvir$ is
given by 
	\begin{equation}
		\frac{d\nh}{d\mvir}=f(\sigma)\frac{\rhom}{M_{\rm vir}^2}\left|\frac{d\ln\sigma^{-1}}{d\ln \mvir}\right|.
		\label{hmf}
	\end{equation}
where $\sigma$ is the amplitude of the perturbations and 
$f(\sigma)$ is the halo multiplicity function. 
The cumulative number density of halos above the mass \mvir\ is simply:
\begin{equation}
\nh(>\mvir) = \int_{M_{\rm vir}}^{\infty}\frac{d\nh}{d\log\mvir}d\log\mvir.
\label{chmf}
\end{equation}
In this paper we use the 
parametrization given in \citet{Tinker+2008}:
	\begin{equation}
		f(\sigma) = A \left[ \left(\frac{\sigma}{b}\right)^{-a} + 1 \right] e^{-c / \sigma^2}.
	\label{Tinker}
	\end{equation}
The amplitude of the perturbations is given by
	\begin{equation}
		\sigma^2(\mvir, a) = \frac{D^2(a)}{2\pi^2}\int P(k) k^2 W^2(k,\mvir) dk,
		\label{sigma_mvir}
	\end{equation}
where $P(k)$ is the power spectrum of perturbations and $W^2(k,\mvir)$ is a 
window function defined to be the Fourier transform of the real-space top-hat filter
of a sphere of mass $\mvir$ and $D(a)$ is the linear growth-rate factor of
the perturbations given by the expression
	\begin{equation}
		D(a) = \frac{g(a)}{g(1)},
	\end{equation}
where to a good approximation $g(a)$ is given by \citep{Lahav+1991}:
	\begin{equation}
		g(a) =\frac{ \frac{5}{2}\Omega_{\rm m}(a)a}{\Omega_{\rm m}(a) - \Omega_{\Lambda}(a) + [1 + \frac{1}{2} \Omega_{\rm m}(a) ] / [1 + \frac{1}{70}  \Omega_{\Lambda}(a) ] }.
	\end{equation}

\begin{figure}
	\vspace*{-130pt}
	\hspace*{-20pt}
	\includegraphics[height=4.5in,width=4.5in]{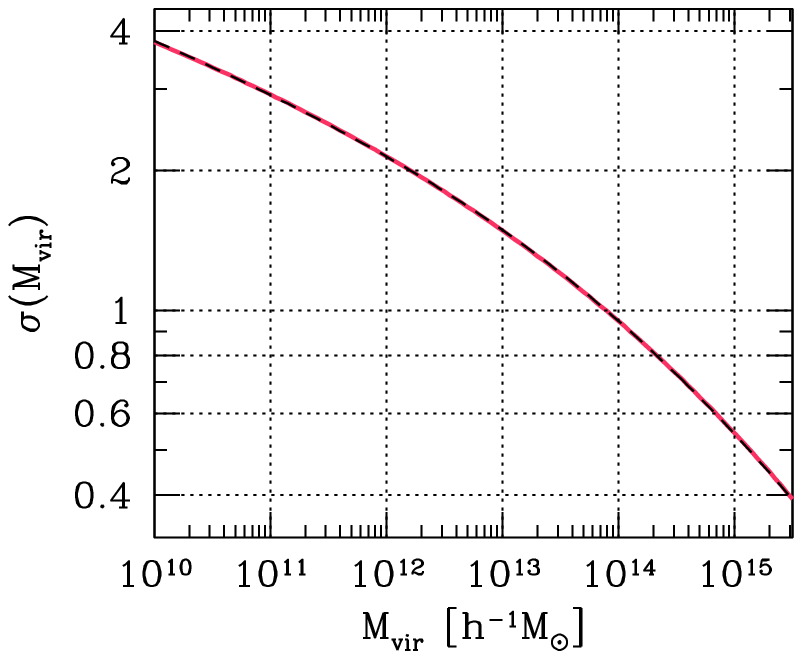}
		\caption{Amplitude of linear perturbations, $\sigma(\mvir)$, as a function
		of \mvir.  The red solid line shows the numerical solution to 
		Equation (\ref{sigma_mvir}). 
		The dashed black line shows the fit to the amplitude of perturbations given
		by Equation (\ref{sigma_mvir_fit}).
 	}
	\label{f6a}
\end{figure}

Figure \ref{f6a}
shows the amplitude of perturbations, $\sigma(\mvir)$, as a function of \mvir\  for $a = 1$. 
The red solid line is calculated based on the transfer function from CAMB \citep{Lewis+2000}. 
Following \citet{Klypin+2011}, we find that the amplitude of perturbations $\sigma$ is well given by 
	\begin{equation}
		\sigma(\mvir) = \frac{17.111y^{0.405}}{1+1.306y^{0.22}+6.218y^{0.317}},
		\label{sigma_mvir_fit}
	\end{equation}
with $y =  1 / \mvirtw$.
Note that the above fit is only valid for the Bolshoi-Planck cosmology studied in this paper. 
The dashed line in Figure \ref{f6a} shows the best fitting model for $\sigma(\mvir)$
given by the above equations.  

\begin{figure}
	\vspace*{-155pt}
	\hspace*{-20pt}
	\includegraphics[height=4.5in,width=4.5in]{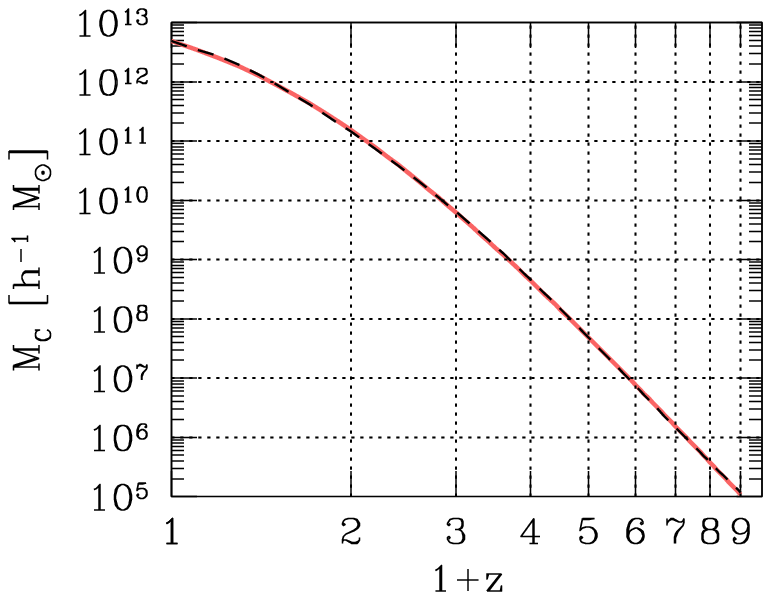}
		\caption{Characteristic halo mass $M_{\rm C}$ as a function of redshift. The red
		solid line shows the numerical solution to Equation (\ref{MPS_equation}).
		The dashed black line shows our numerical fit to $M_{\rm C}$ 
		given by Equation (\ref{sigma_mvir_fit}).
 	}
	\label{fMPS}
\end{figure}

The characteristic mass $M_{\rm C}(z)$ of halos just collapsing at redshift $z$  is given by $\nu=1$ in 
	\begin{equation}
		\nu = \frac{\delta_c}{\sigma(M_{\rm C},z)} .
		\label{MPS_equation}
	\end{equation}
where $\delta_c = 1.686$ for uniform-density spherical collapse. Figure \ref{fMPS} shows the 
Characteristic halo mass $M_{\rm C}$ as a function of redshift, red solid line. Note the
strong dependence with redshift.
For our cosmology, at $z=0$ we find that $M_{\rm C} \sim 5\times10^{12}h^{-1}\msun$ while 
at $z=1, 2$ and 5 we find that $M_{\rm C} \sim 1.5\times10^{11},$  $6.3\times10^{9}$ and $7.6\times10^{8}h^{-1}$ \msun, respectively. 
We find that to good accuracy the redshift dependence of $M_{\rm C}$ is given by the following fitting function,
	\begin{equation}
		\log M_{\rm C} ( a ) = 12.68 - 0.084 y^{0.01} - 5.33 y^{1.92} - 8.22 y^{7.8} ,
	\label{MPS}
	\end{equation}
where $y \equiv z/(1+z) = 1-a$.

Next, we update the best fitting parameters to the \citet{Tinker+2008} formulae for the virial halo mass definition for
the Planck cosmology. 
In order to find the best fitting parameters to the
redshift evolution of the halo mass function we will assume the following redshift
dependence for the parameters $\chi_{i}  = A, a, b$ and $c$ in Equation (\ref{Tinker}), 
\begin{equation}
\chi_{i} = \chi_{0,i} + \chi_{1,i} z + \chi_{2,i} z^2.
\label{TinkerFit}
\end{equation}
 Table \ref{T3} lists the best fit parameters to the redshift evolution of the halo mass function.
In the left panel of Figure \ref{f6} we present the best fits to the resulting halo mass functions from the simulations. 
For completeness, in the right panel of the same Figure we also show the resulting cumulative halo mass function 
using the best fit parameters from table \ref{T3}.

\begin{figure*}
	\vspace*{-190pt}
	\hspace*{-20pt}
	\includegraphics[height=6.5in,width=6.5in]{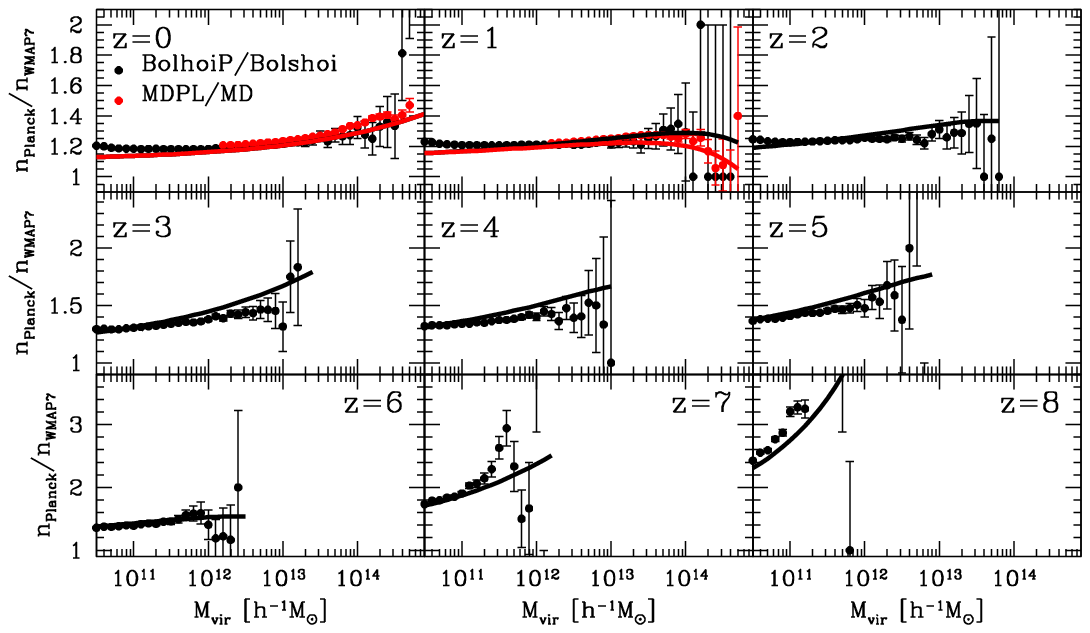}
		\caption{The ratio of the number densities of distinct halos between the 
		Bolshoi-Planck $n_{\rm BP}$ and
		the Bolshoi $n_{\rm B}$ simulations as a function of \mvir\ from $z=0$ to $z=8$ are shown as filled circles, and the ratio of the fitting functions are shown as solid lines.
 	}
	\label{f11}
\end{figure*}

Figure \ref{f11} shows the ratio of the number densities $n_{\rm BP}$ and
$n_{\rm B}$ between the Bolshoi-Planck and the Bolshoi simulations as a function of \mvir\ from $z=0$ to $z=8$.  The different
cosmological parameters imply that at  $z=0$, on average, there are $\sim12\%$ more Milky-Way mass halos in the Bolshoi-Planck
than in the Bolshoi simulation.  This fraction increases to higher masses, $\sim25\%$ for
$\mvir\sim3\times10^{13}h^{-1}\msun$. This fraction also increases with redshift, and we find that at $z=2, 4$ and 6
there are $\sim25,40$ and $60\%$ more Milky-Way mass halos in the Bolshoi-Planck than in the Bolshoi simulation.  At $z=8$, there are about 3 times as many $\mvir = 10^{11}h^{-1} \msun$ halos in Bolshoi-Planck as in Bolshoi.

In the cold dark matter cosmology it is predicted that the number density of dark matter halos is a strong function of
halo mass at low masses $ d\nh / d\mvir \propto M_{\rm vir} ^ {-1.8}$. In contrast, the observed galaxy stellar mass function, 
as well as the luminosity function, has a slope that is flatter. Recent analysis have found slopes between $\alpha\sim1.4-1.6$
\citep{Blanton+2005,Baldry+2008,Baldry+2012} meaning that, for some reason, the 
star formation efficiency in low mass halos has been suppressed \citep[e.g.][]{Behroozi+2013,Moster+2013}. Nevertheless, measurements of the baryonic mass 
have found slopes as steep as $\alpha\sim1.9$ \citep{Baldry+2008}.

\subsection{Subhalo mass function}

\begin{figure*}
	\vspace*{-130pt}
	\hspace*{5pt}
	\includegraphics[height=4in,width=4in]{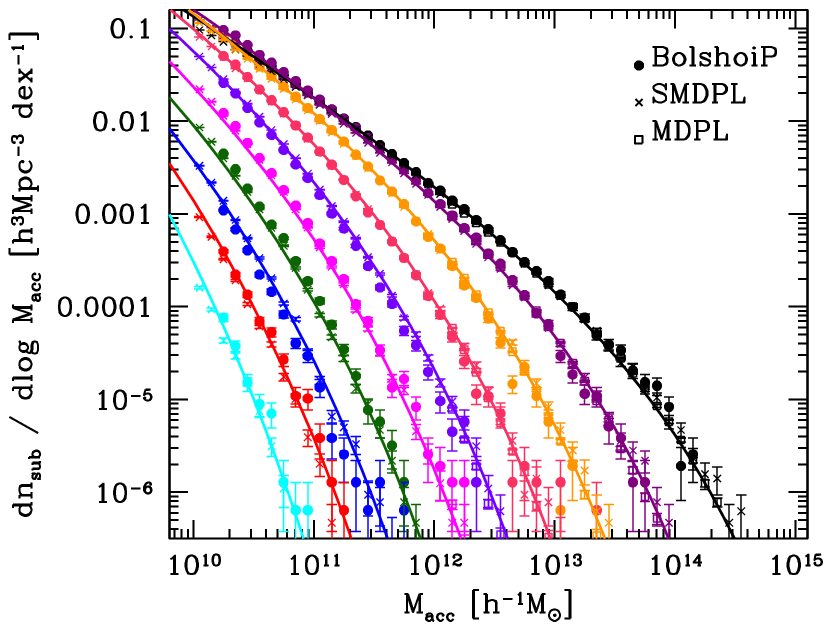}
	\hspace*{-90pt}
	\includegraphics[height=4in,width=4in]{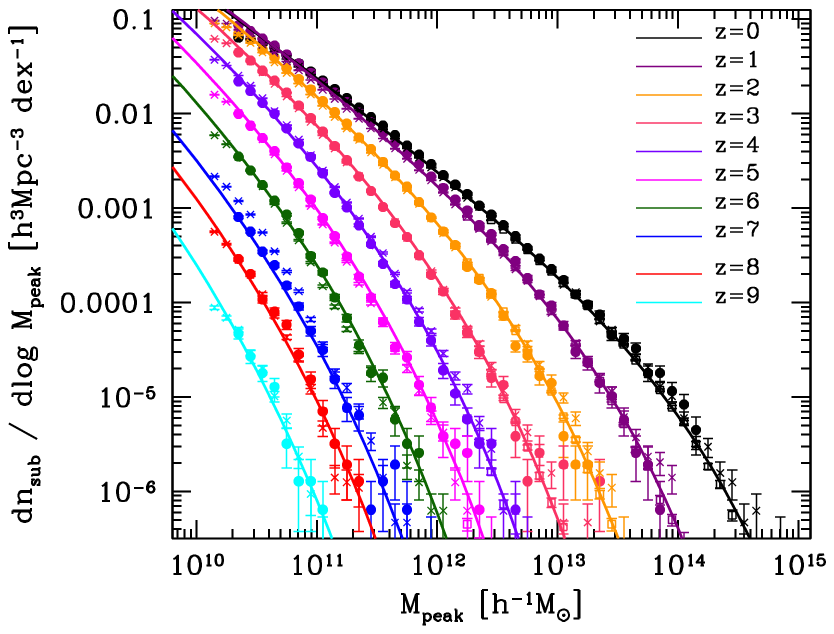}
		\caption{Subhalo mass \macc\  ({\bf left panel}) and \mpeak\ ({\bf right panel}) functions are shown for redshifts from $z=0$ to 9, along with fitting functions Equations (\ref{f16fita}) to (\ref{f16fitc}). 
 	}
	\label{f16}
\end{figure*}

\begin{table}
	\caption{Best fit parameters for the subhalo mass function.}
	\begin{center}
		\begin{tabular}{c c c}
			\hline
			\hline	
			 Parameter & \macc & \mpeak  \\
			\hline
			\hline	
			 $C_0$ & -0.1213 & -0.0863\\
			\hline
			$C_1$  & 0.0113   & 0.0087 \\
			\hline
			$C_2$   & -0.0168  & -0.0113\\
			\hline
			$C_3$    & -0.0032 & -0.0039\\
			\hline
			$C_4$    & 0.0005  & 0.0004\\
			\hline
			$ \alpha_{{\rm sub},1}$  & 0.1810 & 0.0724\\
			\hline
			$ \alpha_{{\rm sub},2}$  & 0.2138 & 0.2206\\
			\hline
			 $M_{0} [h^{-1}\msun]$ & 11.1416 & 11.9046\\
			\hline
			$M_{1} [h^{-1}\msun]$  & -0.6595 & -0.6364 \\
			\hline
			$M_{2} [h^{-1}\msun]$ & -0.0015 & -0.02069\\
			\hline
			$M_{3} [h^{-1}\msun]$  & 0.0183 & 0.0220\\
			\hline
			$M_{4} [h^{-1}\msun]$ & -0.00164  & -0.0012\\
			\hline
			\hline
		\end{tabular}
		\end{center}
		\label{T4}
\end{table}

\begin{figure*}
	\vspace*{-130pt}
	\hspace*{5pt}
	\includegraphics[height=4in,width=4in]{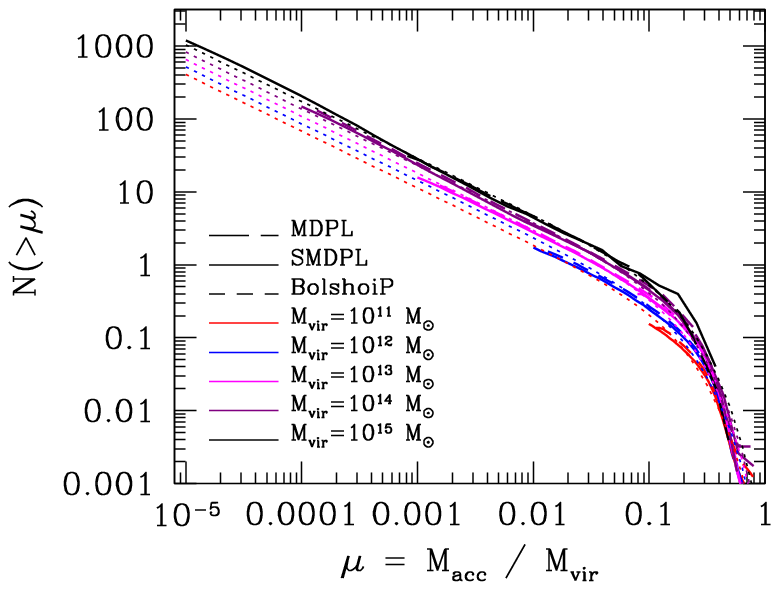}
	\hspace*{-90pt}
	\includegraphics[height=4in,width=4in]{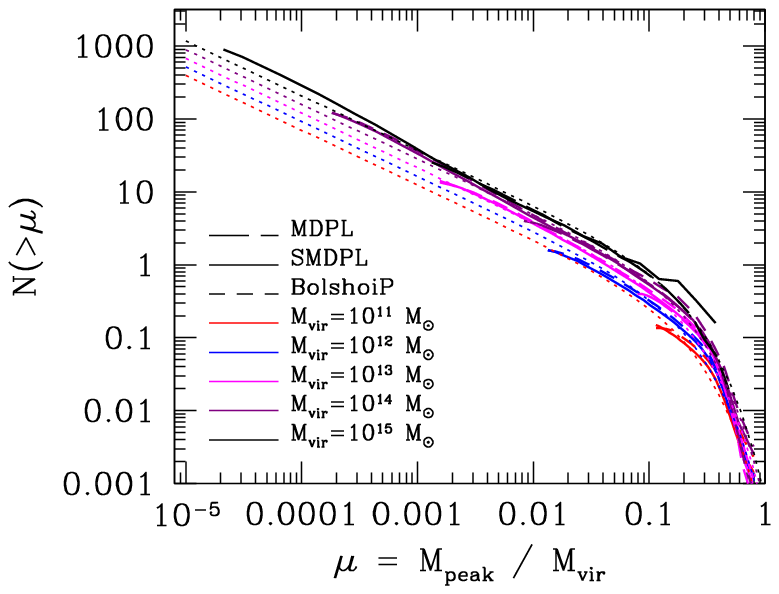}
		\caption{The cumulative subhalo mass function 
		for different host halos as a function of $\mu = \macc / \mvir$ and $\mpeak / \mvir$ 
		is shown as solid curves.  The dotted curve is the fitting function 
		Equation (\ref{nsub_mvir}). 
 	}
	\label{f14}
\end{figure*}

Subhalos can lose a significant fraction of their mass due to tidal 
striping. Since tidal stripping affects the dark matter more than the stars of the central galaxy deep inside the halo, this means that 
the correlation between galaxy stellar mass and present subhalo mass is not trivial. 
Therefore in approaches for connecting galaxies to dark matter (sub)halos, such as the 
abundance matching technique, it has been shown that the mass the subhalo had when it was
still a distinct halo correlates better with the stellar mass of the galaxy it hosts. This comes from the 
fact that when assuming identical stellar-to-halo mass relations for central and satellite galaxies, 
the observed two-point correlation function is reproduced.  Note, however, that while some authors have shown that this is true 
for most of the scales \citep[see e.g.,][]{Conroy+2006, Reddick+2013}, others have argued that
on the very small scales abundance matching fails in reproduce the observed clustering of galaxies \citep[see e.g.,][]{Wang+2007,Gao+2012}.
Therefore it is useful to report the subhalo mass
function when subhalos were accreted for the first time into a bigger distinct halo, i.e., the 
comoving number density of subhalos at the mass range $\log\macc$ and $\log\macc+d\log\macc$.
In addition, we also present results for $\mpeak$ in
the Bolshoi-Planck simulations. Similarly to $\vpeak$, $\mpeak$ is defined as the maximum mass reached 
along the main progenitor assembly mass.  
Figure \ref{f16} shows the redshift evolution of the
subhalo mass function, $d \nsub / d \log \msub$, derived from the Bolshoi-Planck and MD-Planck simulations.
The solid circles in the left panel show the resulting subhalo mass function for \macc\ while the results 
for \mpeak\ are shown in the right panel of the same figure.  

The solid lines in Figure \ref{f16} show our best fitting models to the redshift evolution of the subhalo mass 
function. Previous reports of the mean number of subhalos above some mass $\msub$ 
at a given host of mass \mvir\ 
have found that this is nearly independent of host mass and scales as $\Nsub \propto (\msub / M_{\rm vir})^\alpha$ with
$\alpha\sim-1$, see also below. This implies that to a good approximation the subhalo mass function
is $d \nsub / d \log \msub\propto  M_{\rm vir}^{\alpha} d\nh / d\log\mvir$ \citep[see e.g.,][]{Behroozi+2013}. 
Here we generalized the fitting model proposed in \citet{Behroozi+2013} for the redshift evolution of the subhalo mass function for both \macc\ and \mpeak:
	\begin{equation}
		\frac{d\nsub}{d\log\msub}=C_{\rm sub}(z) G(\mvir,z)\frac{d\nh}{d\log\mvir},
	\end{equation}
where 
	\begin{equation}
		\log C_{\rm sub}(z) = C_0 + C_1 a + C_2 a^2 + C_3 a^3 + C_4 a^4,
	\label{f16fita}
	\end{equation}
and 
	\begin{equation}
	 	 G(\mvir,z) = X^{\alpha_{{\rm sub},1}}  \exp ( - X^{\alpha_{{\rm sub},2} }),
		\label{f16fitb}
	\end{equation}
where $X = \mvir / M_{\rm cut}(z)$. The fitting function for $ M_{\rm cut}(z)$ is
given by
\begin{equation}
\log(M_{\rm cut}(z)) = M_0 + M_1 z + M_2 z^2 + M_3 z^3 + M_4 z^4 .
\label{f16fitc}
\end{equation}

\subsection{Number of subhalos as a function of their host halo mass}

\begin{figure}
	\vspace*{-130pt}
	\hspace*{-20pt}
	\includegraphics[height=4.5in,width=4.5in]{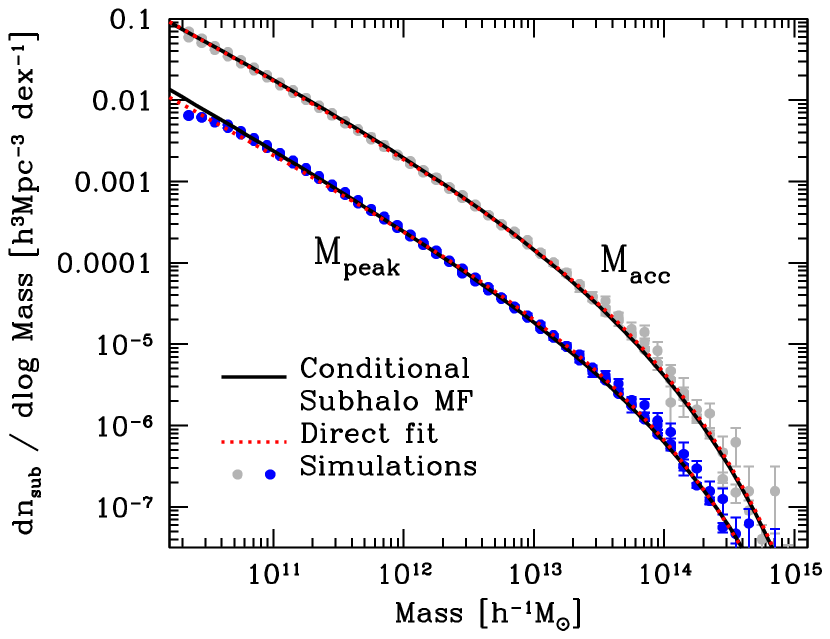}
		\caption{Comparison between the measured subhalo mass function from
the \bpl, \smdpl\ and \mdpl\ simulations and by computing  $d \nsub / d \log \msub$
when using Equation (\ref{nsub_HOD}) for \macc\ and \mpeak.
 	}
	\label{suhmf_comp}
\end{figure}

\begin{figure}
	\vspace*{-20pt}
	\hspace*{-10pt}
	\includegraphics[height=4.7in,width=4.7in]{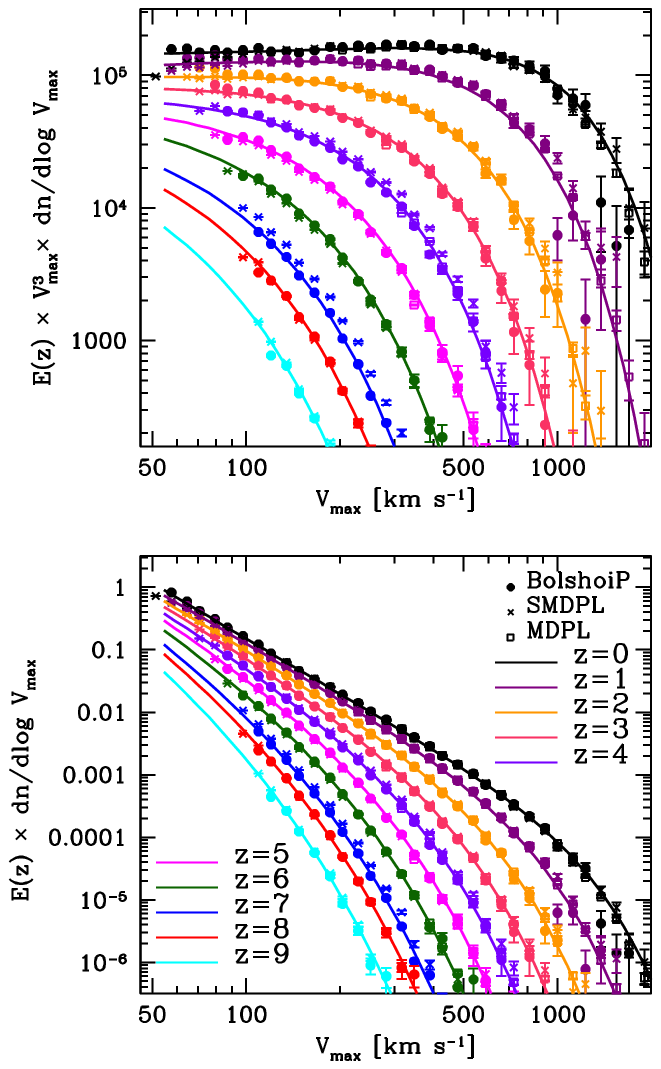}
		\caption{Maximum circular velocity function of distinct halos from $z=0$ to $z=9$. 
		The different solid lines show the fits to the simulation, Equation (\ref{Vmax-fcn}). 
		{\bf Upper Panel:} The product $E(z)\times V^{3}_{\rm max} \times\ d\nh/d\log\vmax$ is
		shown, to split the evolution of the velocity function. {\bf Bottom Panel:} Same 
		but for the product $E(z) \times\ d\nh/d\log\vmax$. 
 	}
	\label{f12}
\end{figure}

Characterizing the number of subhalos in hosts of different masses \mvir\ is relevant for several reasons.  The predicted abundance of satellites in Milky Way mass galaxies has been a very active topic since $N-$body
numerical simulations could resolve subhalos in galactic halos \citep{Klypin+1999,Moore+1999}. 
Subhalos are the natural sites for satellite galaxies, thus, using statistical approaches that connect the stellar mass
of central/satellite galaxies to dark matter halos/subhalos allows prediction of the abundance of satellite galaxies 
as function of the stellar mass of their host. 
Previous studies have used these predictions in order to make direct 
comparisons to what is observed from large galaxy groups catalogues \citep[][and reference therein]{Yang+2009,RDA12,RAD13}. 
These studies have found that in order to
reconcile the observed abundance of satellite galaxies in clusters of different masses, the galaxy stellar-to-halo 
mass relation of central and satellite galaxies should be different, especially at lower masses \citep[see also, ][]{Neistein+2011,Wetzel+2013}. 

Figure \ref{f14} shows the mean cumulative number of subhalos for various host halos with masses
$\mvir=10^{11}, 10^{12}, 10^{13}$, and $10^{14}$\msun, solid lines. In the left panel of this
figure we present the results when defining subhalo mass at the time when they first became subhalos, i.e.,
at their time of first accretion, \macc. The right panel of the same figure presents the results 
when defining the mass of subhalos when they reached the maximum mass over their main branch, \mpeak. 

Following \citet{Boylan-Kolchin+2010}, in this paper we parametrise the mean 
cumulative number of subhalos at a given host halo mass, \mvir,  as
	\begin{equation}
		\Nsub=\mu_0\left(\frac{\mu}{\mu_1}\right)^a\exp\left[-\left(\frac{\mu}{\mu_{\rm cut}}\right)^b\right],
		\label{nsub_mvir}
	\end{equation}
\begin{equation}
 \mu_0 = (\mvirtw)^c
\end{equation}
where $\mu=\msub/\mvir$ and $\mu_0$ is a normalisation term that depends on \mvir. 
We use the above functional form for both \macc\ and \mpeak. For \macc\ we find that 
$a = -0.777$, $b = 1.210$, $\mu_1 = 0.030$, $\mu_{\rm cut}= 0.199$ and $c = 0.102$. For \mpeak\ 
we find that 
$a = -0.749$, $b = 1.088$, $\mu_1 = 0.042$, $\mu_{\rm cut}= 0.199$ and $c = 0.118$.  

It is interesting to see the consistency of Equation (\ref{nsub_mvir}) with the
values reported in the literature based on the spatial clustering of galaxies. 
Previous studies have concluded that the mean occupation number of satellite galaxies above some stellar mass $M_*$ increases roughly proportionally to halo mass, i.e., $N_{\rm sat}\propto \mvir$  \citep[see, e.g.,][]{Zehavi+2005,Zehavi+2011}.  A recent study to redshift $z=1.2$ \citep{Skibba+2015} found $N_{\rm sat}\propto M_{\rm vir}^\alpha$ with favored values of $\alpha\approx 0.95 $ to 1.2.
Ignoring the exponential term in  Equation (\ref{nsub_mvir}) for $\msub \lesssim 0.1\mvir$, we find that mean occupation number of subhalos (above some mass $M_{\rm sub}$) is $N_{\rm sub} \propto M_{\rm vir}^{c-a} \approx M_{\rm vir}^{0.9}$, which is consistent with observations. 

Note that in Equation 
(\ref{nsub_mvir}) the parameter $\mu_1$ gives the typical fractional mass of the most massive 
subhalo relative to the host halo \mvir\ for Milky Way sized halos  (i.e., $\mvir\sim10^{12}h^{-1}\msun$), 
and thus $\mu_1\times\mvir$ gives the typical
mass of the most massive subhalo. When defining subhalo masses at the time of their first
accretion, we find that the typical mass of the most massive subhalo is $\sim3\%$ of its host halo mass.  
For Milky Way mass halos, the most massive subhalo typically
has a mass of $\macc\sim3\times10^{10}h^{-1}\msun$, which corresponds to a stellar mass of
$\sim10^{7.9}\msun$ based on abundance matching results. This is more than
an order of magnitude lower than the LMC. When defining subhalo masses 
as $\mpeak$, we find that the typical mass of the most massive subhalo is $\sim4\%$ of its host halo, which is also more
than an order of magnitude less than the LMC.  Indeed, only a small fraction of Milky Way mass galaxies have a satellite as massive as the LMC \citep{Busha+2011,RP+2013}, which has a total mass $\sim 10^{11} M_\odot$ \citep[e.g.,][]{Besla2015} including its dark matter halo.

Note that the number of subhalos of mass between $\log\msub$ and $\log\msub+d\log\msub$ 
residing in halos of mass $\mvir$, usually referred as the
conditional subhalo mass function, can be obtained by simply differentiating \Nsub
	\begin{equation}
		\Phisub =\frac{d\Nsub}{d\log\msub}.
	\end{equation}
We can therefore use this definition to infer the subhalo mass function:
	\begin{equation}
		\frac{d\nsub}{d\log\msub} = \int \Phisub \frac{d\nh}{d\log\mvir}d\log\mvir.
		\label{nsub_HOD}
	\end{equation}
Figure \ref{suhmf_comp} shows the comparison between the measured subhalo mass function from
the \bpl, \smdpl\ and \mdpl\ simulations and by computing  $d \nsub / d \log \msub$
when using Equation (\ref{nsub_HOD}) both for \macc\ and \mpeak. We find that Equation (\ref{nsub_HOD}) 
provides an accurate prescription for the  subhalo mass function.

\section{Halo and subhalo velocity function}
\begin{table}
	\caption{Best fit parameters for Equation (\ref{vmax_param}).}
	\begin{center}
		\begin{tabular}{c c c c c c}
			\hline
			\hline	
			$\chi_{i}$ & $\chi_{0,i}$ & $\chi_{1,i}$ &  $\chi_{2,i}$ & ${\alpha_1}$ & ${\alpha_2}$ \\
			\hline
			\hline
			 $\log (A / E(z))$ & 4.785 & -0.207 & 0.011 & 0.897 & 1.856\\
			\hline
			$a$  & -1.120  & 0.394 & 0.306 & 0.081 & 0.554 \\
			\hline
			$b$  & 1.883 & -0.146 & 0.005  & 1 & 2\\
			\hline
			$\log V_0$  & 2.941 & -0.169 & 0.002 & 1 & 2\\
			\hline
			\hline
		\end{tabular}
		\end{center}
	\label{T5}
\end{table}

\begin{figure*}
	\vspace*{-250pt}
	\hspace*{-20pt}
	\includegraphics[height=6.5in,width=6.5in]{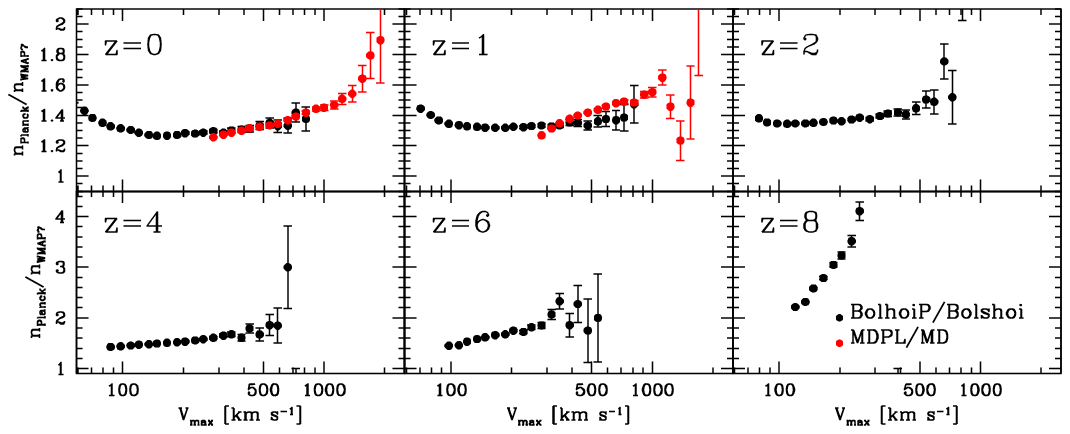}
		\caption{The ratio of the distinct halo number densites between the 
		Bolshoi-Planck $n_{\rm BP}$ and the Bolshoi $n_{\rm B}$ simulations as a 
		function of \vmax\ at $z=0, 1,2,4,6$ and $z=8$.  
 	}
	\label{f10}
\end{figure*}

The comoving number density of distinct halos with maximum circular velocity between 
$\log\vmax$ and $\log\vmax+d\log\vmax$---i.e., the maximum circular velocity function $(d\nh/d\log\vmax)$---is recognized as a sensitive probe of dark matter \citep{ColeKaiser1989,Shimasaku1993,Gonzalez+2000,
Zavala+2009,Papastergis+2010,Trujillo-Gomez+2011,Schneider+2014,Klypin+2015,Papastergis+2015}. 
Figure \ref{f12} shows the maximum circular velocity function $d\nh/d\log\vmax$ 
from $z=0$ to $z=8$ for the Bolshoi-Planck and MD-Planck simulations. 
The upper panel of this figure shows the 
product $E(z)\times V^{3}_{\rm vmax} \times\ d\nh/d\log\vmax$ while the bottom 
panel shows the product $E(z) \times\ d\nh/d\log\vmax$. Recall that $E(z)$ is
the expansion rate, Equation (\ref{E(z)}). 

In this paper we parametrize the velocity function as
	\begin{equation}
	\frac{d\nh}{d\vmax} = A V_{\rm max}^{-2}  \left[ \left(\frac{\vmax}{V_0}\right)^{-a} + 1 \right]
		\exp\left[{-\left(\frac{\vmax}{V_0}\right)^{b}} \right] .
	\label{Vmax-fcn}
	\end{equation}
We assume the following redshift
dependence for the parameters $\chi_{i}  = \log(A / E(z)) , a, b$ and $\log V_0$ 
	\begin{equation}
		\chi_{i} = \chi_{0,i} + \chi_{1,i} z^{\alpha_1} + \chi_{2,i} z^{\alpha_2}.
		\label{vmax_param}
	\end{equation}
Table \ref{T5} lists the best fit parameters for the velocity function. 

Figure \ref{f10} shows the ratio of the number densities of distinct halos between the Bolshoi-Planck, $n_{\rm BP}$, and
the Bolshoi $n_{\rm B}$ simulations as a function of \vmax\ at $z=0, 1,2,4,6$ and $z=8$.  At 
$z=0$, on average, there are $\sim25\%$ more halos with $\vmax = 200$ km/s in the Bolshoi-Planck
than in the Bolshoi simulation, and this stays practically constant for most $\vmax$ values up
to $z=2$. This fraction increases at $z=4,6$ and more drastically at $z=8$ where we find that 
there are $\sim60,78$ and $258\%$ more $\vmax = 200$ km/s halos in the Bolshoi-Planck
than in the Bolshoi simulation. 

\begin{figure}
	\vspace*{-20pt}
	\hspace*{-10pt}
	\includegraphics[height=4.5in,width=4.5in]{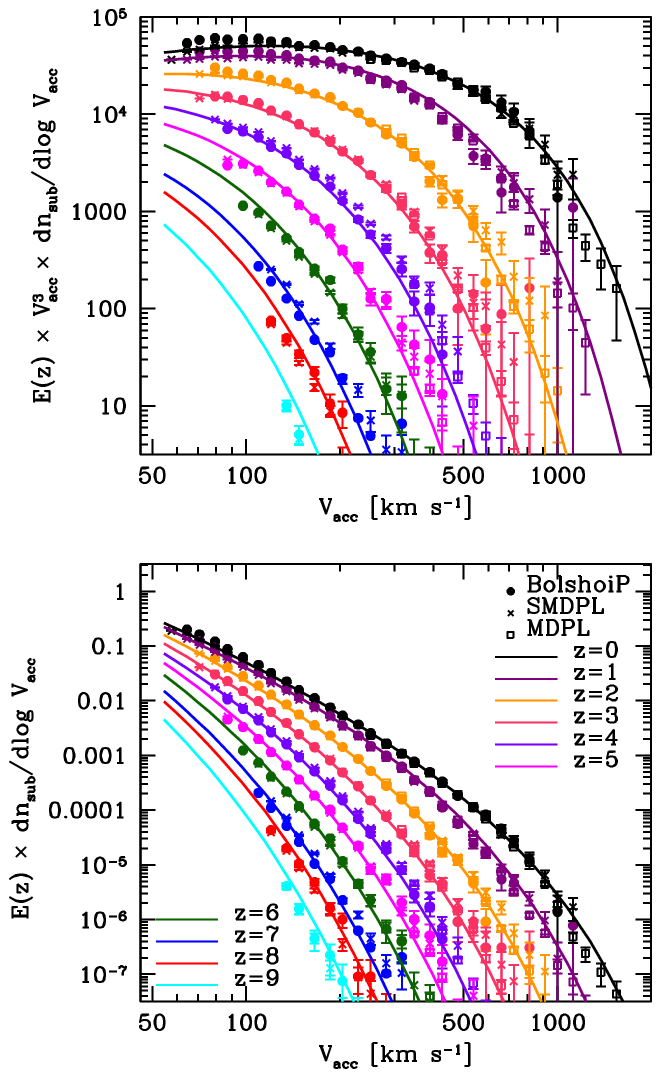}
		\caption{Redshift evolution of the subhalo maximum circular velocity function, with circular velocity $V_{\rm acc}$ measured at accretion.
 	}
	\label{f17a}
\end{figure}

\begin{figure}
	\vspace*{-20pt}
	\hspace*{-10pt}
	\includegraphics[height=4.5in,width=4.5in]{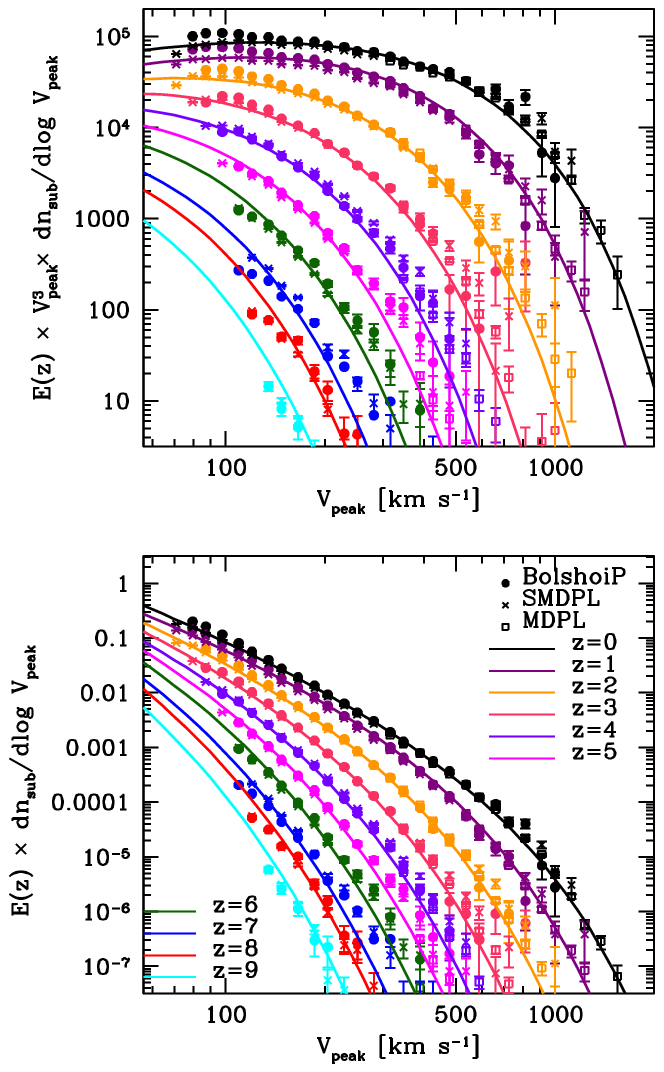}
		\caption{Redshift evolution of the subhalo circular velocity function, as a function of the subhalo's peak circular velocity $V_{\rm peak}$.
 	}
	\label{f17}
\end{figure}

\begin{table}
	\caption{Best fit parameters to the subhalo 
maximum circular velocity function. }
	\begin{center}
		\begin{tabular}{c c c}
			\hline
			\hline	
			 Parameter & \vacc & \vpeak  \\
			\hline
			\hline	
			 $C_0$ & -0.6768 & -0.5800 \\
			\hline
			$C_1$ & 1.3098 & 1.5905  \\
			\hline
			$C_2$  & -1.1288 &-1.1360  \\
			\hline
			$C_3$ &  0.0090  & -0.0378 \\
			\hline
			$C_4$ &  0.214820  & 0.18092 \\
			\hline
			$ \alpha_{{\rm sub},1}$ & 1.1375 & 1.1583  \\
			\hline
			$ \alpha_{{\rm sub},2}$ & 0.5200  & 0.5806 \\
			\hline
			 $V_{0}$ [km s$^{-1}$] & 0.2595 & 0.5410  \\
			\hline
			$V_{1}$ [km s$^{-1}$] & 3.5144 & 3.4335  \\
			\hline
			$V_{2}$ [km s$^{-1}$] & -2.8817 & -3.0026  \\
			\hline
			$V_{3}$ [km s$^{-1}$] & -0.3910 & -0.3687 \\
			\hline
			$V_{4}$ [km s$^{-1}$] & 0.8729 & 0.9450 \\
			\hline
			\hline
		\end{tabular}
		\end{center}
		\label{T6}
\end{table}

Figure \ref{f17a} shows the redshift evolution of the subhalo 
maximum circular velocity function, $d\nsub / d\log\vsub$, at the time at the 
time of their accretion, \vacc. Similarly, Figure \ref{f17} shows $d\nsub / d\log\vsub$
for \vpeak. Recall that  $\vpeak$ is defined as the maximum circular velocity reached 
along the main progenitor assembly. \citet{Reddick+2013} found that \vpeak\ is a better proxy for galaxy stellar mass/luminosity than alternatives such as \vacc\ or \mpeak. 
The solid lines in Figures \ref{f17a} and  \ref{f17} show the best fits to $d\nsub / d\log\vsub$. 
We motivate the fitting functional form for $d\nsub / d\log\vsub$ using the same arguments
as for $d\nsub / d\log\msub$. As we will show later, the mean number of subhalos above some
maximum circular velocity  is nearly independent of host maximum velocity. To be specific, we 
parameterize $d\nsub / d\log\vsub$ using the following functional form:
	\begin{equation}
		\frac{d\nsub}{d\log\vsub}=C_{\rm sub}(z) G(\vsub,z)\frac{d\nh}{d\log\vmax},
	\label{Vsubfcn}
	\end{equation}
where 
	\begin{equation}
		\log C_{\rm sub}(z) = C_0 + C_1 a + C_2 a^2 + C_3 a^3 + C_4 a^4,
	\end{equation}
and 
	\begin{equation}
	 	 G(\mvir,z) = X^{\alpha_{{\rm sub},1}}  \exp ( - X^{\alpha_{{\rm sub},2} }),
	\end{equation}
where $X = \vmax / V_{\rm cut}(z)$.  The function fit form for $ V_{\rm cut}(z)$ is
given by 
\begin{equation}
\log(V_{\rm cut}(z)) = V_0 + V_1 z +
V_2 z^2 + V_3 z^3 + V_4 z^4.
\end{equation}

\begin{figure*}
	\vspace*{-130pt}
	\hspace*{5pt}
	\includegraphics[height=4in,width=4in]{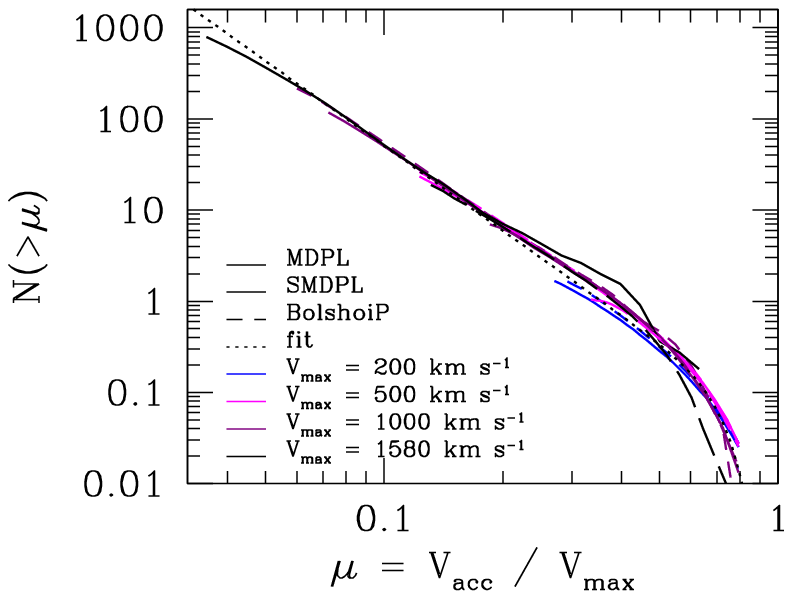}
	\hspace*{-90pt}
	\includegraphics[height=4in,width=4in]{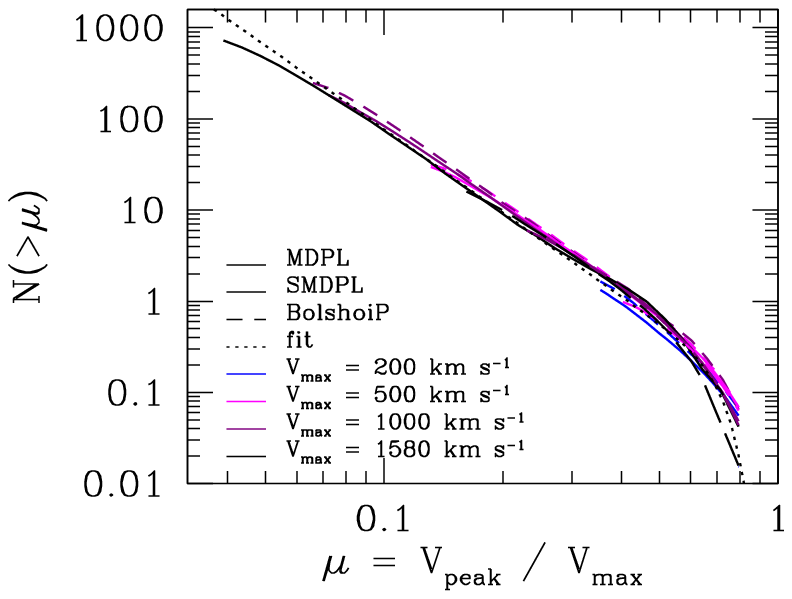}
		\caption{Mean cumulative number of subhalos of maximum circular velocity $V_{\rm sub}$ for host halos with
		maximum velocities $\vmax = 200, 500, 1000$ and $1580$ km /s as a function of $V_{\rm sub}/V_{\rm max}$ for  ({\bf left panel}) $V_{\rm sub} = V_{\rm acc}$, and  ({\bf right panel}) $V_{\rm sub} = V_{\rm peak}$.  The dotted curve is the fitting function Equation (\ref{vsub_mvir}).
 	}
	\label{f15}
\end{figure*}

Figure \ref{f15} shows the mean cumulative number of subhalos of maximum circular velocity \vsub\ 
for various host halos with maximum circular velocities 
$\vmax=200, 500, 1000$ and $\vmax = 1580$ km/s for the \bpl, \smdpl\ and \mdpl\ simulations (dashed, solid and long dashed lines respectively). 
In the left panel of this
figure we present the results when defining $\vsub = \vacc$. 
The right panel of the same figure presents the results 
when defining $\vsub = \vpeak$. Analogously to the mean 
cumulative number of subhalos at a given host halo mass, \Nsub, 
we parameterize the mean cumulative number of subhalos above some 
maximum circular velocity at a given host halo maximum circular velocity as:
	\begin{equation}
		\Nvsub=\left(\frac{\mu}{\mu_1}\right)^a\exp\left[-\left(\frac{\mu}{\mu_{\rm cut}}\right)^b\right],
		\label{vsub_mvir}
	\end{equation}
where $\mu=\vsub/\vmax$. We use the above functional form for both \vacc\ and \vpeak. 
The resulting best fitting parameters when using \vacc\ are 
$a = -3.0881$, $b = 7.545$, $\mu_1 = 0.356$ and $\mu_{\rm cut}= 0.736$, while for
\vpeak\ are 
$a = -3.045$, $b = 8.850$, $\mu_1 = 0.416$ and $\mu_{\rm cut}= 0.738$.
Similarly to Equation 
(\ref{nsub_mvir}), the parameter $\mu_1$ in Equation (\ref{vsub_mvir}) 
gives the typical fractional maximum circular velocity of the most  
massive subhalo relative to the host halo. For host halos with velocities of $\sim200$ km/s (MW-sized halo)
the typical most massive subhalo has a velocity of $\vacc\sim 71.2$ km/s and of $\vpeak\sim 83.2$ km/s.
These values are more consistent with the velocity of the LMC. See also \citet{Busha+2011} which compared 
the number of satellites as massive as the LMC/SMC in the Bolshoi simulation with the number observed in
MW-size galaxies in SDSS.

The number of subhalos of maximum circular velocity between $\log\vsub$ and $\log\vsub+d\log\vsub$ 
residing in halos of maximum circular velocity $\vmax$, referred as the
conditional subhalo maximum circular velocity function, is given by
\begin{equation}
\Vhisub =\frac{d\Nvsub}{d\log\vsub}.
\end{equation}
Using this definition we can thus derive the maximum circular velocity function as:
\begin{equation}
\frac{d\nsub}{d\log\vsub} = \int \Vhisub \frac{d\nh}{d\log\vmax}d\log\vmax.
\label{nvmax_HOD}
\end{equation}

\section{Halo concentration and spin}
\label{C-lambda}

\subsection{Halo concentrations} 
\label{concentrations}

\begin{figure*}
	\vspace*{-250pt}
	\hspace*{-20pt}
	\includegraphics[height=6.5in,width=6.5in]{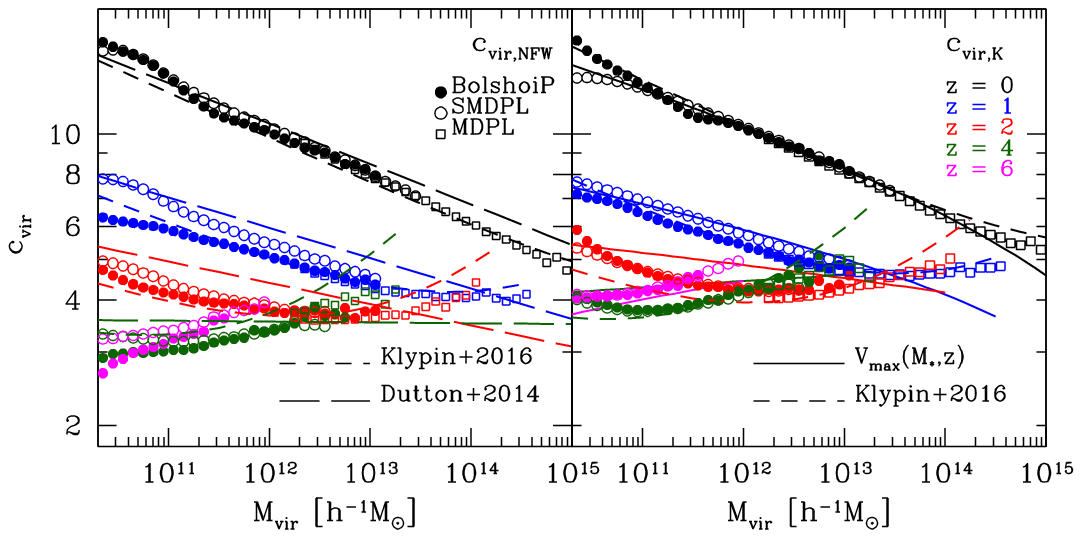}
		\caption{Halo concentration as a function of \mvir\ at $z = 0,1, 2, 4,$ and $6$.
		The left panel in the figure shows halo concentrations calculated by finding the
		scale radius, $R_s$ assuming a NFW profile in the simulation.  For 
		comparison we added the halo concentration-mass relations reported in
		\citet{Klypin+2014} for all halos selected by mass and from \citet{Dutton+2014} for relaxed halos. 
		The right panel
		shows Klypin halo concentrations from determining the scale radius, $R_s$ using the
		$\vmax$ and $\mvir$ relationship from the NFW formulae (see text).
		Solid lines in the left panel show the resulting Klypin 
		concentrations by solving Equation (\ref{RsKlypin}) and using the best fitting
		values for the $\vmax-\mvir$ relation from Section \ref{vmax_section}.  For 
		comparison we added the halo concentration-mass relations reported in
		\citet{Klypin+2014} for all halos selected by \vmax.
 	}
	\label{f3}
\end{figure*}

High resolution $N-$body simulations have shown that the density profile of dark matter halos can be well described 
by the \citet[][NFW]{NFW96} profile,
\begin{equation}
\rho_{\rm NFW}(r) = \frac{4\rho_s}{(r/R_s)(1+r/R_s)^2} .
\end{equation}
The scale radius $R_s$ is the radius where the logarithmic slope of the density profile is -2. The NFW profile is completely characterized by two parameters, for example $\rho_s$ and $R_s$, or more usefully the halo mass, \mvir, and its
concentration parameter, \cvir. The concentration parameter is defined as the ratio between the virial radius $R_{\rm vir}$ and
the scale radius $R_s$:
\begin{equation}
\cvir = \frac{R_{\rm vir}}{R_s} .
\end{equation}

Figure \ref{f3} shows halo concentrations, \cvir, as a function of \mvir\ for redshifts $z = 0,1, 2, 4,$ and $6$. 
The left panel of the figure shows halo concentrations calculated by finding the
best scale radius, $R_s$ assuming a NFW profile for each halo in the simulation. Instead, the right panel
shows halo concentrations calculated by determining the scale radius, $R_s$ using the
$\vmax$ and $\mvir$ relationship from the NFW formulae, see \citet{Klypin+2011} and   \citet{Klypin+2014};  \citep[see also,][]{ROCKSTAR}.  For the NFW profile, the radius at which the circular velocity is maximized is $R_{\rm max} = 2.1626 R_s$ \citep{Klypin+2001,ROCKSTAR}, and it can be shown that 
 \begin{equation}
\frac{\cvir}{f(\cvir)} = V_{\rm max}^2 \frac{R_{\rm vir}}{G M_{\rm vir}} \frac{2.1626}{f(2.1626)} 
\label{RsKlypin}
\end{equation}
where
\begin{equation}
f(x) \equiv \ln(1+x) - \frac{x}{1+x} .
\end{equation}
The Klypin concentration $c_{\rm vir,K}$ can be found be solving Equation (\ref{RsKlypin}) numerically.  It is more robust than determining $R_s$ by fitting the NFW profile, especially for halos with few particles, since halo profiles are not well determined both at distances comparable to the simulation force resolution and also at large distances near $R_{\rm vir}$.
Figure \ref{f3} shows that at high redshifts NFW concentrations are systematically lower than Klypin concentrations.   Fitting functions for $c_{\rm vir,K}$ are given in \citet{Klypin+2014} for all halos and for relaxed halos, for both Bolshoi-Planck/MultiDark-Planck and Bolshoi/MultiDark simulations; fitting functions are also given there for concentrations of halos defined by the 200c overdensity criterion.  Key processes that drive the evolution of halo concentration are also discussed there.  \citet{DiemerKravtsov2015} discusses the relation between halo concentration, the slope of the fluctuation power spectrum and the peak height.

The solid lines in the left panel of Figure \ref{f3} show the resulting Klypin 
concentrations by solving Equation (\ref{RsKlypin}) and using the best fitting
values for the $\vmax-\mvir$ relation from Section \ref{vmax_section}, see Equation (\ref{vmax-mh}). 
At $z = 0$ and $z=1$ the resulting concentrations are in very good agreement with 
what is found in the simulation with an accuracy of $\sim3\%$ for halos above 
$\mvir = 10^{10}h^{-1}\msun$. However, at higher redshifts $z =2,4,6$, our predicted Klypin 
concentrations have an accuracy of $\sim10\%$. 

\subsection{Halo Spin} 

\begin{figure*}
	\vspace*{-250pt}
	\hspace*{-20pt}
	\includegraphics[height=6.5in,width=6.5in]{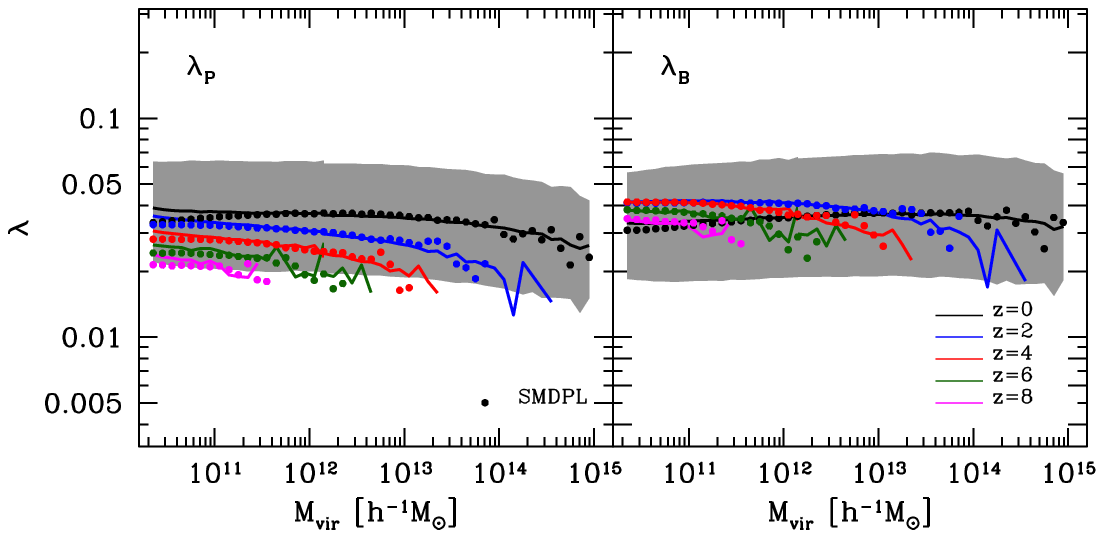}
		\caption{Spin parameter as a function of \mvir\ at $z = 0, 2, 4,6,$ and $8$. Medians are shown as the solid lines. At
		$z = 0$ the grey area is the $68\%$ range of the distribution. 
		The left panel of this figure shows the spin parameter calculated using Equation (\ref{lambdaP}) while the right panel shows the spin parameter calculated using Equation (\ref{lambdaB}).
 	}
	\label{f3a}
\end{figure*}

\begin{figure*}
	\vspace*{-150pt}
	\hspace*{-20pt}
	\includegraphics[height=6.5in,width=6.5in]{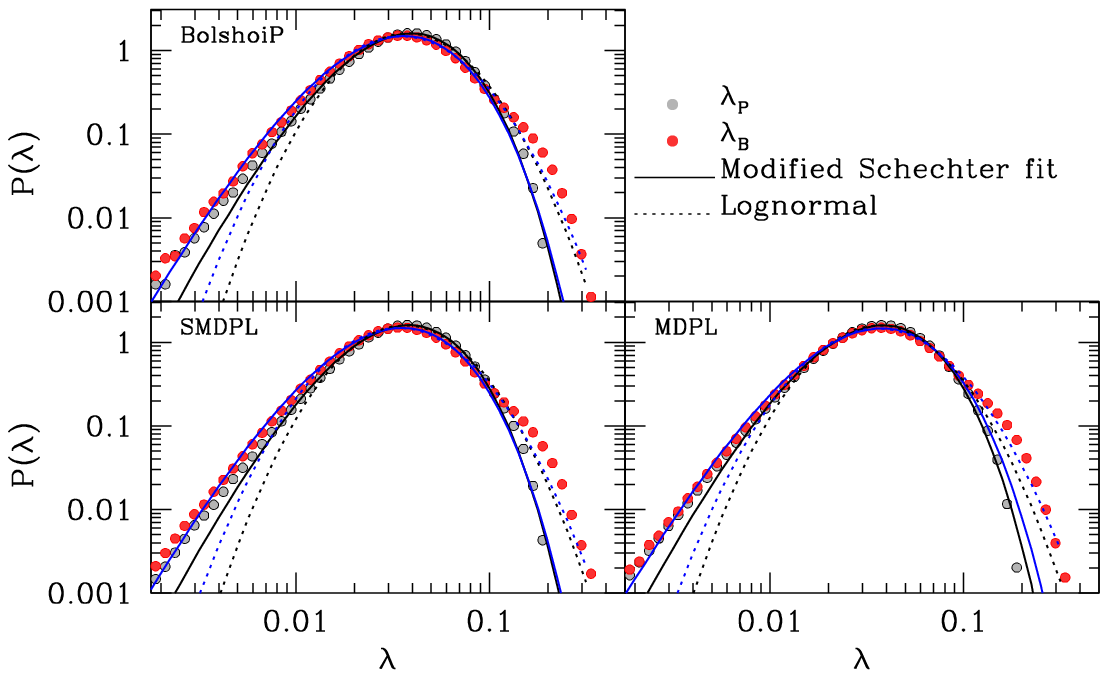}
	\vspace*{-30pt}
	\caption{Halo spin distribution for the \bpl\ (upper left), \smdpl\ (bottom left) and \mdpl\ (bottom right) 
	simulations.  Filled grey circles show the $\lambda_{P}$ distribution while the red circles
	show the $\lambda_{B}$ distribution. Solid lines show the best fit to a Schechter-like function, Equation (\ref{Sch}),
	while the dotted lines show the best fit for a lognormal distribution, Equation (\ref{logn}). 
	The black (blue) lines are the best fits to the $\lambda_{B}$ ($\lambda_{P}$) distributions. 	
	Both distributions are well fit at low values by the Schechter-like distribution, which also is a 	
	good fit the the $\lambda_{P}$ distribution at higher values, while  $\lambda_{B}$ is 			
	somewhat better fit by a log-normal distribution at higher values.  }
\label{lambda_distrb}
\end{figure*}

\begin{table}
	\caption{Best fit parameters to the lognormal distribution function $P(\log\lambda)d\log\lambda$.}
	\begin{center}
		\begin{tabular}{c c c c c}
			\hline
			\hline	
			Simulation & $\sigma_{\rm P}$ & $\log\lambda_{\rm 0, P}$ & $\sigma_{\rm B}$ & $\log\lambda_{\rm 0, B}$  \\
			\hline
			\hline
			BolshoiP & 0.248 & -1.423 & 0.268 & -1.459 \\
			\hline
			SMDPL  & 0.249 & -1.435 & 0.268 & -1.471 \\
			\hline
			MDPL  & 0.250 & --1.438 & 0.271 & --1.443 \\
			\hline
			\hline
		\end{tabular}
		\end{center}
	\label{Tspin1}
\end{table}

\begin{table*}
	\caption{Best fit parameters to Schechter-like distribution function for $P(\log\lambda)d\log\lambda$.}
	\begin{center}
		\begin{tabular}{c c c c c c c}
			\hline
			\hline	
			Simulation & $\alpha_{\rm P}$ & $\beta_{\rm P}$ & $\log\lambda_{\rm 0, P}$ & $\alpha_{\rm B}$ & $\beta_{\rm B}$ &  $\log\lambda_{\rm 0, B}$  \\
			\hline
			\hline
			BolshoiP & 4.126 & 0.610 & -2.919 & 3.488 & 0.6042 & -2.878 \\
			\hline
			SMDPL  & 4.090 & 0.612 & -2.917 & 4.121 & 0.611 & -2.916 \\
			\hline
			MDPL  & 4.047 & 0.612 & -2.914 & 3.468 & 0.591 & -2.907 \\
			\hline
			\hline
		\end{tabular}
		\end{center}
	\label{Tspin2}
\end{table*}

The left panel of Figure \ref{f3a} shows the medians for the spin parameter $\lambda_{\rm P}$ as a function of \mvir\ 
at $z = 0, 2, 4,6,$ and $8$. The spin parameter for every halo in the simulations was calculated using the 
definition \citep{Peebles1968}:
	\begin{equation}
		\lambda_{\rm P} = \frac{J  | E |^{1/2}} { G M_{\rm vir}^{5/2}},
	\label{lambdaP}
	\end{equation}
where $J$ and $E$ are the total angular momentum and the total energy of a halo of mass \mvir. As others have found, 
the spin parameter  $\lambda_{\rm P}$ correlates only weakly with halo mass especially at $z = 0$. 
The median value for Milky Way mass halos (i.e., with $\mvir \sim 10^{12} h^{-1}\msun$) at $z = 0$ is $\lambda_{\rm P} \sim0.036$, and it decreases a factor 
of $\sim1.8$ at $z=6$, that is, $\lambda_{\rm P} \sim0.02$. For Milky Way mass halos, the dispersion is approximately 
$\sim 0.24$ dex at $z = 0$ and it decreases to $\sim 0.16$ dex at $z=6$. Note that the dispersion is not symmetric, meaning
that the distribution of $\lambda_{\rm P}$ is not a lognormal distribution. This is consistent with previous findings 
based on high resolution $N-$body simulations \citep[e.g.,][]{Bett+2007}. 

 The right panel of Figure \ref{f3a} shows the spin distribution calculated using the alternative 
definition \citep{Bullock+2001}: 
\begin{equation}
\lambda_{\rm B} = \frac{J}{\sqrt{2} M_{\rm vir} V_{\rm vir} R_{\rm vir}},
\label{lambdaB}
\end{equation}
which can be obtained from Equation (\ref{lambdaP}) by assuming all particles to be in circular orbits. 
Similarly to $\lambda_{\rm P}$, the spin parameter  $\lambda_{\rm B}$ correlates only weakly with halo mass especially at $z = 0$. We found that the median value for Milky 
Way mass halos at $z = 0$ is $\lambda_{\rm P} \sim0.035$ and it decreases to $\lambda_{\rm P} \sim0.027$ at $z = 6$. 
For Milky Way mass halos, the dispersion of  $\lambda_{\rm B}$ is slightly larger than of $\lambda_{\rm P}$; we find that  it is
$\sim 0.27$ dex at $z = 0$ and it decreases to $\sim0.2$ dex at $z=6$.

The spin parameter  $\lambda_{\rm B}$ slightly increases at high redshifts especially for low mass halos, $\mvir\lesssim10^{12}\msun$.
In contrast, the value of the spin parameter $\lambda_{\rm P}$ shows a systematic decrease as redshift increases.  This was previously noted over the interval $z=0-2$ by \citet{HetzneckerBurkert}, who attribute the different evolution of the two spin parameters mainly to different effects of minor mergers on $\lambda_{\rm P}$ and $\lambda_{\rm B}$.

Figure \ref{lambda_distrb} quantifies in more detail the distribution of halo spins separately for the \bpl, \smdpl\ and \mdpl\
simulations. In order to avoid resolution effects and to obtain reliable statistics, we calculate the distribution of halo spins
in the halo mass range $10^{11}-10^{14}h^{-1}\msun$ for the \bpl\ (upper left panel) and \smdpl\ (bottom left panel) 
simulations, while for the \mdpl\ (bottom right panel) simulation
we do the same but for the mass range $10^{12}-10^{14}h^{-1}\msun$. In all the panels the grey filled circles show 
the distribution for $\lambda_{\rm P}$ while the red filled circles show the distribution for $\lambda_{\rm B}$. As anticipated from
the $\lambda-\mvir$ relationship, the $\log \lambda$ distributions are asymmetrical. This is more evident for $\lambda_{\rm P}$
than for $\lambda_{\rm B}$. In order to quantify this we try to fit all the distributions using a lognormal probability
distribution:
\begin{equation}
P(\log\lambda) = \frac{1}{\sqrt{2\pi\sigma^2}}\exp\left(-\frac{\log^2(\lambda/\lambda_0)}{2\sigma^2}\right).
\label{logn}
\end{equation}
The best fit parameters of $P(\log\lambda)$ both for $\lambda_{P}$ and for $\lambda_{B}$ are listed in Table \ref{Tspin1}. 
We find that while the lognormal distribution gives a fairly good description for $P(\log\lambda_{B})$ this is not the case for $P(\log\lambda_{P})$
for all the simulations. In particular, the distribution has too many halos with low values of $\lambda_{P}$. 
In order to provide a more accurate description of the halo distribution we propose to use a Schechter-like function given by 
\begin{equation}
P(\log\lambda) = A f(\lambda),
\label{Sch}
\end{equation}
where
\begin{equation}
f(\lambda) = \left(\frac{\lambda} {\lambda_0}\right)^{-\alpha} \exp\left[-\left(\frac{\lambda} {\lambda_0}\right)^\beta\right],
\end{equation}
\begin{equation}
A = \left[\int _{-\infty}^{\infty} f(\lambda) d\lambda \right]^{-1}. 
\end{equation}
The best fit parameters for \bpl, \smdpl\ and \mdpl\ simulations for both $\lambda_{P}$ and  
$\lambda_{B}$ are listed in Table \ref{Tspin1} for the log-normal distribution and Table \ref{Tspin2} 
for the Schechter-like distribution.
We find that a Schechter-like function gives a more accurate prescription for the distribution of 
$\lambda_{\rm P}$ than for $\lambda_{\rm B}$. In particular, this distribution has some problems in reproducing the tail of
high $\lambda_{\rm B}$ that declines more like a lognormal distribution.

It is thought that the angular momentum of galaxies is related the angular momentum of dark matter halos and thus to their 
spin parameter. Under this assumption, the scale length of disk galaxies, $R_{\rm d}$, can be obtained in terms of $\lambda$ and
$R_{\rm vir}$. Specifically the relation is given by $R_{\rm d} \propto \lambda \times R_{\rm vir}\propto \lambda \times M_{\rm vir}^{1/3}$. As before, if we assume for simplicity
that the $M_b/\mvir$ ratio is constant, the relation between a galaxy's radius and its baryonic mass is given by
$R_{\rm d}\propto \lambda \times M_b^{1/3}$. Note that the scatter of the size--mass relation is just the resulting
scatter of the $\lambda-\mvir$ relation. Indeed, 
the dispersion of the spin parameter, either $\lambda_{\rm P}$  or $\lambda_{\rm B}$, is very similar 
to the observed dispersion of disk galaxy scale lengths at least at low redshifts where reliable measurements can be obtained 
\citep[see e.g.,][]{Mosleh+2013}. 

Note that the different redshift evolution of the $\lambda_{\rm P}-\mvir$ and $\lambda_{\rm B}-\mvir$ relations leads to different predictions
of the $R_{\rm d} - M_{\rm b}$ relation and its evolution. In particular, models of galaxy formation calculating galaxy sizes based on the spin
parameter $\lambda_{\rm B}$ will result in more extend galaxies (and potentially in larger numbers of low surface brightness galaxies)
at high redshifts compared to those models using $\lambda_{\rm P}$.  Is also possible that galaxy star formation rates could be affected since more extended galaxies 
presumably have lower gas surface densities than more compact disks, and thus lower SFRs according to the Kennicutt-Schmidt law. 

Two recent papers have discussed the evolution of galaxy sizes out to redshift $z \sim 8$ using Hubble Space Telescope images, mainly from the CANDELS survey.  \citet{Shibuya+2015} finds that the median effective radius $r_e$ evolves with redshift as $r_e \propto (1+z)^{-1.3}$, with no evolution in the slope, the median S\'ersic index ($n \sim 1.5$), or the standard deviation of the log-normal distribution.  They find that the ratio of the effective radius to the virial radius of the halos is nearly constant at $r_e/R_{\rm vir} = 0.01 - 0.035$.  This is just what one would expect from the lack of redshift evolution in $\lambda_{\rm B}$, while the factor of $\sim 2$ decline in 
$\lambda_{\rm P}$ from $z=0$ to 8 would predict a corresponding decline in the ratio $r_e/R_{\rm vir}$.  The other recent paper, \citet{Curtis-Lake+2014}, finds a slower decline of effective radius with redshift, and in fact cannot reject the possibility that there is no size evolution.  This is possibly consistent with the modest increase with redshift of $\lambda_{\rm B}$ for lower mass halos, and inconsistent with the expected decrease in $r_e/R_{\rm vir}$ from the decline in 
$\lambda_{\rm P}$.  The radii of these high-redshift galaxies are being measured in rest-frame UV, which is typically rather clumpy \citep{Shibuya+2016,Curtis-Lake+2014}.  It will be very interesting to see what sort of galaxy size evolution with higher redshifts is revealed by James Webb Space Telescope at rest-frame optical wavelengths.

\section{On the evolution of $\vmax-M_*$}
\begin{figure}
	\vspace*{-150pt}
	\hspace*{-20pt}
	\includegraphics[height=4.5in,width=4.5in]{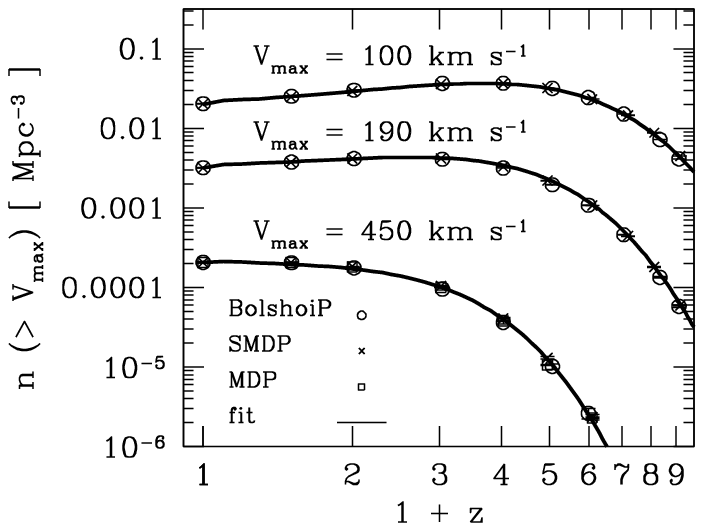}
		\caption{Evolution of the velocity function \vmax\ for fixed $\vmax = 100, 190$ and $450$ km /s
		for the \bpl, \smdpl, and \mdpl\ simulations. The solid lines are the fits to Equation (\ref{Vmax-fcn}).
		For low velocities the velocity function is practically constant after redshift $z\sim4$, while for
		high velocity halos it is nearly constant after redshift $z\sim1$. }
	\label{Vmax_cte}
\end{figure}

\begin{figure}
	\vspace*{-210pt}
	\hspace*{-40pt}
	\includegraphics[height=5.5in,width=5.8in]{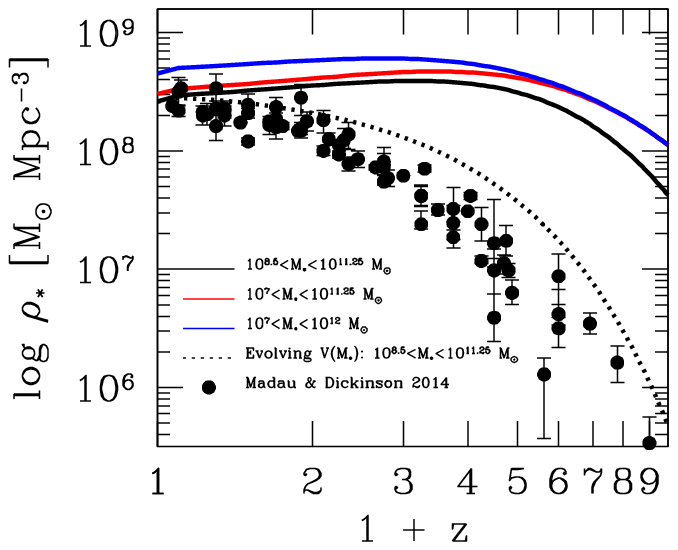}
		\caption{Cosmic stellar mass density since $z\sim9$.  Filled circles show the observations compiled in \citet{MadauDickinson2014}.  The solid curves show the predicted cosmic stellar mass density using fits to the Tully-Fisher and Faber-Jackson velocity-to-stellar-mass relations as described in the text, assuming that these relations are independent of redshift and that the \vmax\ of dark matter halos is the same as the $V_{\rm max,g}$ of galaxies. The solid black line
shows the predicted cosmic stellar mass density for a range of stellar masses $\log(M_\ast/M_\odot) = 8.5 - 11.25$. Comparing the red curve, for  $\log(M_\ast/M_\odot) = 7 - 11.25$, to the black curve shows that including lower stellar masses increases $\rho_\ast$ more at high redshifts; comparing the blue curve, for  $\log(M_\ast/M_\odot) = 7 - 12$, to the black and red curves shows that   including higher stellar masses increases $\rho_\ast$ more at low redshifts.
Clearly, all of these predictions are inconsistent
with observations at $z\grtsim 1$---they produce too much stellar mass density at early redshifts, and the wider stellar mass range represented by the blue curve exceeds the observed stellar mass density even at $z=0$. 
Since the stellar mass function is evolving, velocity-mass relations like Tully-Fisher must also evolve. 
The dotted lines show the predictions when using a model in which
the maximum circular velocity-to-stellar mass relation evolves with redshift
as described in the text. 
 	}
	\label{TFfig}
\end{figure}

Early determinations of the evolution in the maximum circular velocity and the stellar mass/luminosity relations---the Tully-Fisher relation for spiral galaxies and the Faber-Jackson relation for ellipticals---have found only a weak evolution from $z\sim0.85$ to $z\sim0$ \citep{Conselice+2005}. This result has been further supported and generalized in \citet{Kassin+2007} from $z\sim1$ to $z\sim0$,
based on fairly large samples of galaxies from AEGIS and DEEP2 and adopting the indicator
 $S_{0.5} ^2=  0.5 V^2_{\rm max} + \sigma_g^2$ which accounts for disordered motions 
 \citep{Weiner+2006,Covington+2010}. On the other hand, observations indicate that at $z \lesssim 1.5$ the number density of star forming galaxies at a fixed velocity evolves very little, while 
the number density of quiescent galaxies evolves more rapidly \citep[e.g.,][]{Bezanson+2012}. 

 Figure \ref{Vmax_cte} shows that the comoving number density of low circular velocity halos is nearly constant since $z\sim4$
 but high mass/velocity halos have more evolution. Halos of a given circular velocity at high redshift are lower in mass but denser than halos of the same circular velocity at lower redshift.  To what extent is the nearly constant comoving number density of halos as a function of their circular velocity
 consistent with the weak evolution of the Tully-Fisher relation? This is particularly interesting if the galaxy stellar mass function evolves with 
 redshift, as was first pointed out in \citet{Bullock+2001b}. 
 
 In this section we investigate the above question assuming that the 
 Tully-Fisher and Faber-Jackson relations do not evolve with redshift and that 
 there is a one-to-one correspondence between the maximum circular velocity of
 halos and galaxies, i.e., $\vmax = V_{\rm max,g}$. To do so, our first step is to convert  the 
 Tully-Fisher and Faber-Jackson relations into circular velocities. Arguments based 
 on the Jeans equation in virialized systems result in the relation $V_c = K \sigma$, where
 typical values for $K$ are $\sqrt{2}-\sqrt{3}$ \citep{BinneyTremaine1987}. While there is
 an extensive discussion in the literature of what is the right value for $K$, following \citet{Dutton+2011}
here we assume that $K=1.54$ which is a value halfway between different groups. The next 
step is to derive an average maximum circular velocity-to-stellar mass relation for all galaxies:
$\langle\log V_{\rm max,g} \rangle= \langle\log V_{\rm max,g}\rangle(\log M_*)$.  The method
is to use the average Tully-Fisher and Faber-Jackson relations for local galaxies and take into 
account the observed fraction of disk and elliptical galaxies. For simplicity, we assume that 
all disk galaxies are star-forming systems while ellipticals correspond to quiescent galaxies. 
Then the average maximum circular velocity is given by
\begin{equation}
 \langle\log V_{\rm max,g}\rangle =f_{\rm SF}  \langle\log V_{\rm max,TF}\rangle +  f_{\rm Q}\langle\log V_{\rm max,FJ}\rangle.
 \label{Vmax_gal}
\end{equation}
where $\langle\log V_{\rm max,FJ}\rangle = \langle  \log(1.54\sigma)\rangle$ and $f_{\rm SF} =1 - f_{\rm Q} $. 
Note that the above equation depends on stellar mass.  We take
the fraction of quiescent galaxies $f_{\rm Q}$ from \citet{Behroozi+2013}, and 
we use the fits for the Tully-Fisher and Faber-Jackson relations reported in \citet{Dutton+2011}. 
The fraction of quiescent galaxies $f_{\rm Q}$ has been taken from \citet{Behroozi+2013}.
We assume for simplicity that the maximum circular velocity of dark matter halos,  \vmax , corresponds to the maximum
circular velocity of galaxies, $V_{\rm max,g}$, i.e., $\vmax = V_{\rm max,g}$. In this way, we can then 
solve Equation (\ref{Vmax_gal}), $\langle\log V_{\rm max,g} \rangle= \langle\log V_{\rm max,g}\rangle(\log M_*)$, 
for $M_*$ and thus transform velocities into stellar mass. 

The black solid line in
Figure \ref{TFfig} shows the predicted cosmic stellar mass density for galaxies with stellar mass $M_{*} = 10^{8.5} - 10^{11.25} \msun$
since $z=9$ assuming that $\langle\log V_{\rm max,g}\rangle(\log M_*)$ is independent of redshift, and the red and blue curves are the same for wider ranges of stellar masses. 
For comparison, we plot a recent compilation from \citet{MadauDickinson2014} of the evolution of the observed stellar mass density. Our simple model represented by the black curve seems to be roughly
consistent with the observational evidence of the weak evolution of the maximum circular velocity since $z\sim1$. In contrast,
at high redshifts the model produces far too many stars. We thus conclude that 
a strong evolution of the Tully-Fisher and Faber-Jackson relations is required to higher redshifts in order to reconcile 
the predicted cosmic stellar mass density with observations. 

Based only on theoretical arguments, is it possible to derive a simple model for the redshift evolution for
$\langle\log V_{\rm max,g}\rangle(\log M_*)$? In Section 3.1, we found
that the evolution of the maximum circular velocity of dark matter halos is well described by
$\vmax\ \propto\ \left[ \mvir E(z)\right]^{\alpha}$. In other words, the zero point of the maximum circular velocity
evolves with $E(z)^{\alpha}$. If we adopt the same reasoning, we can assume that the zero point of the
Tully-Fisher and Faber-Jackson relations evolve with redshift as $E(z)^{\alpha_{\rm TF}}$ and $E(z)^{\alpha_{\rm FJ}}$ 
respectively, where $\alpha_{\rm TF}=0.259$ and $\alpha_{\rm FJ}=0.37$ are their corresponding slopes at $z=0$. 
The dotted line in Figure \ref{TFfig} shows the predicted cosmic star formation rate density based on this simple
evolutionary model. Despite the simplicity of this model, 
the predictions are much more consistent with observations at high redshifts
than the non-evolving model that led to the solid black line in the figure. 
Nevertheless, the above models are very simple and they ignore the fact that the \vmax\ of dark matter halos
is not the $V_{\rm max,g}$ of galaxies, as we discuss in the next section.


\section{Observed Velocity Function of Nearby Galaxies}    
  \label{LocalVolume}
\begin{figure}
	\vspace*{-75pt}
	\hspace*{-20pt}
	\includegraphics[height=4.5in,width=4.5in]{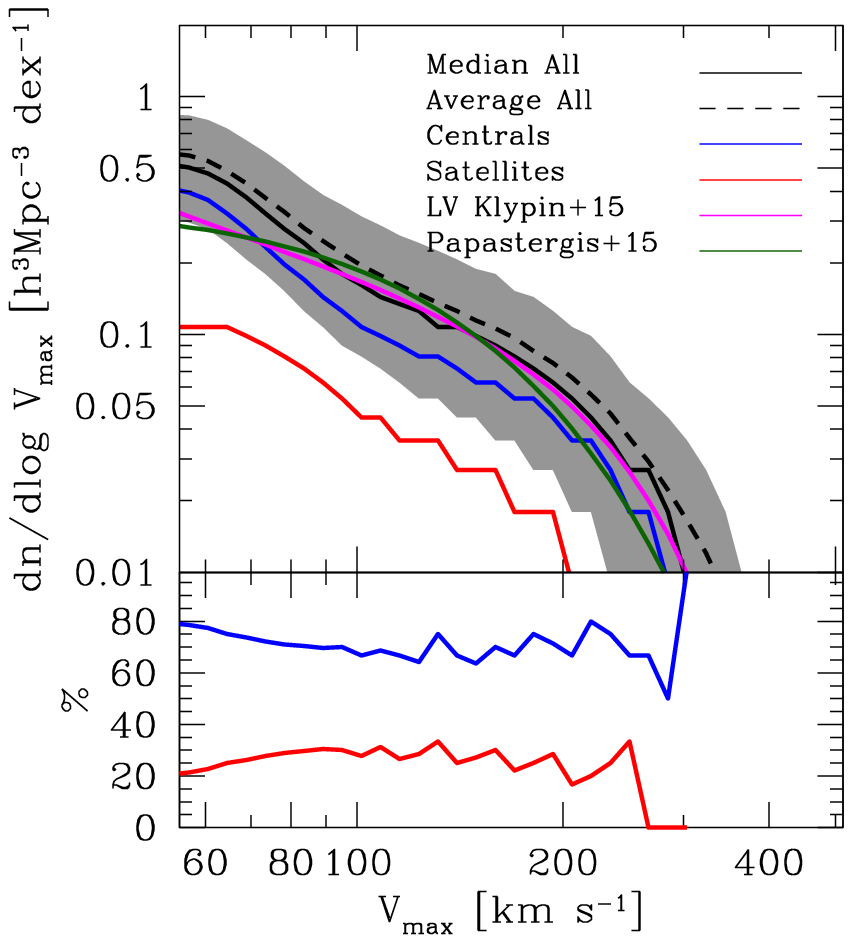}
		\caption{Comparison of the predicted Local Volume 3D velocity function $dN/d \log V$ from the \bpl\ simulation with the observed Local Volume optical velocity function of galaxies within $\sim10$ Mpc \citep[Figure 12 of][]{Klypin+2015} and the HI radio velocity function from the ALFALFA survey \citep{Papastergis+2015}.  }
	\label{V-function}
\end{figure}

Previous studies have considered that
the observed distribution of galaxy velocities is a strong test for galaxy formation models and cosmology 
\citep{ColeKaiser1989,Shimasaku1993,Klypin+1999}. The reason is simply because
the comparison, at a first order, between the theoretical 
halo+subhalo velocity function and the observed velocity function of galaxies
is more direct than the stellar mass/luminosity and halo+subhalo mass functions.
In this section, we compare the local volume galaxy velocity function derived from optical galaxy observations in \citet{Klypin+2015} and
the HI radio galaxy velocity function based on the ALFALFA survey from \citet{Papastergis+2015} to compare
with the theoretical halo+subhalo velocity function from $\Lambda$CDM with the Planck cosmological parameters.

In the past, a number of works have studied the velocity function of halos+subhalos from high resolution N-body simulations
to conclude that it differs from the observed galaxy velocity function by overpredicting the number of low velocities objects. Actually,
these differences are not surprising since a more careful comparison between the theoretical halo+subhalo and galaxy
velocity distributions (1) must include the effects of the baryons on the velocity profile of the halo/subhalo, and 
(2) consistently compare the radii at which galaxies and halo/subhalo velocities (usually \vmax) are measured.  
Indeed, including the effects of the baryons are important since they could increase the maximum circular velocity as a result 
of their gravitational effect over the halo \citep[but see][]{Dutton+2007}. For item (2) we note that local disk galaxies (which are the most extended
objects in the local universe) typically have scale lengths between $R_{\rm d}\sim1- 10$ kpc in the stellar mass range
$10^9-10^{11.5}\msun$. This would imply that $\vmax$ of galaxies would be
observed between $R(\vmax)\sim2- 20$ kpc, for the ideal disk $R(\vmax)\sim2.2\times R_{\rm d}$. 
This is actually markedly different from dark matter halos. Assuming that all halos follow a
NFW profile,  their maximum rotational velocity is reached at $\sim2.16\times R_s$. Based on the
concentrations obtained in Section \ref{concentrations}, this would imply that \vmax\ is reached between $R(\vmax)\sim10-300$ kpc for halos
between $10^{11}-10^{14}$\msun. Clearly using \vmax\ of halos would result in an overestimation of the 
true maximum circular velocity of the galaxy. Although a more proper modeling of these effects requires sophisticated structural and dynamical models of galaxies 
like those described in recent papers \citep{Dutton+2007,Trujillo-Gomez+2011,Dutton+2012,Harrison+2015}, in this paper we follow a more 
empirical approach based on abundance matching. Our goal is to derive a correlation between the maximum circular velocity of galaxies, 
$V_{\rm max,g}$, and dark matter halo/subhalos $\vmax$ without further modeling of galaxy formation. 

\vspace{5pt}
We summarize our algorithm as follows:
\begin{enumerate}
\item To each halo/subhalo in the \bpl\ simulation a galaxy with stellar mass $M_{*}$ is assigned randomly from the
probability distribution function $P(M_{*}|\vpeak)$. This probability distribution function is assumed to be lognormal with mean 
$\langle\log M_{*} \rangle= \langle\log M_{*} \rangle(\log \vpeak)$ obtained from abundance matching 
with halo/subhalo property \vpeak. The scatter of the distribution is assumed to be constant with \vpeak\ with a value of $0.15$ dex. 

\item The next step is to define $P_V(V_{\rm max,g}|M_*)$ as the lognormal probability distribution that a galaxy of mass $M_{*}$ has an observed 
velocity of $V_{\rm max,g}$. The mean of this distribution, $\langle\log V_{\rm max,g} \rangle= \langle\log V_{\rm max,g}\rangle(\log M_*)$, is given by 
 Equation (\ref{Vmax_gal}). We asume that the scatter of the distribution is constant with $M_{*}$ with a value of $0.08$ dex. 

\item  $V_{\rm max,g}$ assignment is based on the argument that at a fixed $M_{*}$, larger $\vpeak$ corresponds to larger $V_{\rm max,g}$. 
This assumption is reasonable since the halo contributes to the total velocity of the galaxy, $V_{\rm max,g}$.
More formally, we obtain galaxy velocities by solving the following equation for $V_{\rm max,g}$ for a given $M_*$:
\begin{equation}
\int^\infty_{V_{\rm max,g}} P_V(V_{\rm max,g}'|M_*) dV_{\rm max,g}' = \int^\infty_{V_{\rm peak}} P(V_{\rm peak}'|M_*)dV_{\rm peak}'.
\end{equation}
\end{enumerate}
In the last equation $P(V_{\rm peak}|M_*)$ is the inverse of the distribution function $P(M_{*}|\vpeak)$.
A few comments are necessary here: In (i) we use a galaxy stellar mass function that has been corrected for
low surface brightness incompleteness in the SDSS and measured over the range between 
$\log M_{*} / \msun =10^{7.6}-10^{12.2}$ (Rodr\'iguez-Puebla et al. in prep). In (ii) we assume that the mean
relation given by  Equation (\ref{Vmax_gal}) is valid in the same mass regime. 
The above procedure ensures that each halo/subhalo in the \bpl\ simulation will host a galaxy
with stellar mass $M_*$ and galaxy velocity $V_{\rm max,g}$.   Recall that the \bpl\ simulation is complete for halos with $V_{\rm max} \grtsim 50$ km/s. 

The Local Volume is a sample of galaxies in a sphere of $\sim 10$ Mpc centered on the Milky Way. \citet{Klypin+2015}
showed that the luminosity function of the Local Volume is consistent with the luminosity function of the local
galaxies in the SDSS. This reflects the fact that the Local Volume is not biased to extreme environments such as voids 
or clusters.
In order to define Local Volume analogs in the \bpl, we have not carried out an exhaustive search to 
find similar environments to the Local Volume. Instead, Local Volume analogs are selected by centering
spheres of 10 Mpc on galaxies with stellar masses in the bin $\log (M_{*} /\msun)\in[10.64,10.84]$ corresponding
to the mass of the Milky-Way \citep{Flynn+2006}, where we have used abundance matching to assign stellar masses to halos in \bpl. 

The upper panel of Figure \ref{V-function} shows the predicted velocity function. The black solid line in the figure shows the median 
velocity function from all the Local Volume analogs found in the simulation, while the dashed curve shows
the average. In order to get a sense of the most common configurations of the Local Volume under our definition, we also present the $68 \%$ range
of the distribution as the grey shaded area. The width of the distribution as a function of mass 
at low velocities is of the order of $\sim0.25$ dex. 
For comparison, the green solid line shows the fitting function to the observed Local Volume velocity function reported in \citet{Klypin+2014}. 
Taking the fairly large dispersion into account, we see that
the predicted velocity function is consistent with observations for galaxies above $\sim 50$ km/s. 
We also find agreement when comparing with the HI radio galaxy velocity function based 
on the ALFALFA survey from \citet{Papastergis+2015}.  In the same figure we show predictions of the velocity 
function decomposed into centrals and satellites. The bottom panel in the same Figure \ref{V-function}  shows
the contributions from centrals and satellites. We find that approximately 
$\sim80\%$ of the galaxies in the Local Volume are centrals.  

\section{Summary \&\ Discussion}

This paper presents many results, both graphically and with fitting functions, from the Bolshoi-Planck and MultiDark-Planck simulations of the large scale structure of the universe, based on the Planck cosmological parameters summarized in Table 1.  Figure \ref{s8-OmM.pdf} shows the WMAP5/7/9 and Planck constraints on the key cosmological parameters $\sigma_8$ and $\Omega_{\rm M}$, and the values adopted for these parameters in many cosmological simulations. 

The Bolshoi-Planck and MultiDark-Planck simulations have been analyzed using \rockstar\ and \ctrees\ to identify and characterize all dark matter halos in all stored time steps, and to construct merger trees of these halos.  In this paper we use the virial radius $R_{\rm vir}$ and virial mass $M_{\rm vir}$, Equation (\ref{Mvir}), to describe these dark matter halos.

It is useful to characterize dark matter halos by their maximum or peak circular velocity, in addition to their virial mass.  Figure \ref{f1} shows the relations between $V_{\rm max}$ and $M_{\rm vir}$ and between  $V_{\rm peak}$ and $M_{\rm vir}$.  The slopes are roughly given by $V \propto M_{\rm vir}^{1/3}$, and we give accurate fitting functions in Equation (\ref{vmax-mh}).

Cosmological simulations also allow determination of the mass accretion rates of the halos as a function of their virial mass and redshift.  Figure \ref{f2} shows these accretion rates three ways: Instantaneous accretion rates (i.e., between stored time steps), accretion rates averaged over the halo's dynamical time, Equation (\ref{tdyn}), and accretion rates of the maximum mass $M_{\rm peak}$ along the main progenitor branch.  We give power-law fitting functions in the form of Equation (\ref{dmhdt-mh}) for all of these, and  a better double-power-law fit, Equation (\ref{dmhdt-mh-dbl}), for the instantaneous accretion rates.  

\citet{RP16} uses abundance matching with the Bolshoi-Planck simulation to explore the possibility that the halo mass accretion rate plays a large part in determining the star formation rate of the central galaxy in the halo, at least when this galaxy lies on the main sequence of star formation.  This Stellar-Halo Accretion Rate Coevolution (SHARC) assumption predicts star formation rates on the main sequence that are in remarkably good agreement with observations for the redshift range $z=0$ to 4.
The paper also shows that the $\sim 0.3$ dex dispersion in the halo mass accretion rates leads to similar small dispersions in the predicted star formation rates, in rough agreement with observations.

Figures \ref{f4} and \ref{f5} show the median halo mass growth and the maximum circular velocity growth as redshift decreases, and also the dispersions of these.  It is remarkable how little change there is with redshift in the circular velocity of halos of $M_{\rm vir}= 10^{11} M_\odot$ at low redshifts.

The \rockstar\ analysis of the Bolshoi-Planck simulation reliably finds all dark matter halos and subhalos with $\grtsim 100$ particles, which corresponds to $M_{\rm vir} \grtsim 2\times10^{10} h^{-1}\msun$ or $V_{\rm max} \grtsim 50$ km/s.  All such halos are expected to host visible galaxies, so the number densities of dark matter halos and subhalos as a function of mass and redshift predict the corresponding abundances of central and satellite galaxies.  We find that the \citet{Tinker+2008} formula Equation (\ref{Tinker}) is a good approximation to the number density of distinct dark matter halos, shown in Figure \ref{f6}, and we give a fitting function Equation (\ref{sigma_mvir_fit}) for the amplitude of fluctuations, shown in Figure \ref{f6a}.  

The characteristic mass $M_{\rm C}(z)$ of $1\sigma$ halos just collapsing at redshift $z$ separates abundant halos with $M_{\rm vir} < M_{\rm C}$ from relatively rare halos with $M_{\rm vir} > M_{\rm C}$.
The clustering properties of halos are also different below and above $M_{\rm C}$ \citep[e.g.,][]{Wechsler+2006}, and halo properties such as their triaxial shapes scale with $M_{\rm C}$ \citep[e.g.,][]{Allgood+2006}.  We plot $M_{\rm C}(z)$ in Figure \ref{fMPS} and give a fitting function Equation (\ref{MPS}) for the Bolshoi-Planck cosmological parameters. 

The maximum circular velocity of dark matter halos is related to observable internal velocities of their central galaxies or the relative velocities of galaxies in groups and clusters, so it is useful to determine the velocity function of distinct halos.  This is plotted in Figure \ref{f12}, with a corresponding fitting function Equation (\ref{Vmax-fcn}).  

The Planck cosmological parameters, especially the higher $\Omega_{\rm M}$ compared with WMAP5/7, result in a greater abundance of halos especially at high masses and redshifts.  This is shown in Figure \ref{f11}.  At $z=0$ there are $\sim12\%$ more $10^{12} h^{-1}\msun$ halos in the Bolshoi-Planck
than in the Bolshoi simulation, and $\sim25\%$ more for $\mvir\sim3\times10^{13}h^{-1}\msun$.  At $z=8$ there are about 3 times as many $\mvir = 10^{11}h^{-1} \msun$ halos in Bolshoi-Planck as in Bolshoi.  Similarly, there are more dark matter halos as a function of $V_{\rm max}$ with the Planck parameters, as shown in Figure \ref{f10}.   At $z=0$ to 2 there are $\sim25\%$ more halos with $\vmax = 200$ km/s in the Bolshoi-Planck
than in the Bolshoi simulation. This fraction increases at $z=4,6$ and 8, with $\sim60,78$ and $258\%$ more $\vmax = 200$ km/s halos in the Bolshoi-Planck simulation. 

Figure \ref{f16}  characterizes the abundance of subhalos as a function of the mass they had at accretion, and also as a function of their peak mass along their major progenitor track, with corresponding fitting function given by Equation (\ref{f16fitb}).  We also plot the redshift evolution of the subhalo maximum circular velocity at accretion in Figure \ref{f17a} and the subhalo peak circular velocity function in Figure \ref{f17}, with fitting functions  Equation (\ref{Vsubfcn}).

It is also useful to know the number of subhalos with a given accreted mass or peak mass compared to the virial mass of the host halo.  This is shown in Figure \ref{f14}, with fitting function Equation (\ref{nsub_mvir}).  It is also useful to know the corresponding numbers of subhalos characterized by their circular velocities.  Figure \ref{f15} shows the number of subhalos as a function of $V_{\rm sub}$ divided by the maximum circular velocity of the host halo, for $V_{\rm sub}$ equal to the subhalo's velocity at accretion $V_{\rm acc}$ or its peak circular velocity $V_{\rm peak}$, with fitting function Equation (\ref{vsub_mvir}).

We calculate the concentration of dark matter halos two ways, by finding the best scale radius $R_s$ assuming a NFW profile or by using Equation (\ref{RsKlypin}).  Figure \ref{f3} shows the resulting concentrations as a function of $M_{\rm vir}$ and redshift from the Bolshoi-Planck, SmallMultiDark-Planck (SMDPL), and MultiDark-Planck (MDPL) simulations.

The halo spin parameter $\lambda$ is a dimensionless way of characterizing the angular momentum of each dark matter halo.  We calculate the halo spin parameter using both the \citet{Peebles1968} definition Equation (\ref{lambdaP}) and the \citet{Bullock+2001} definition Equation (\ref{lambdaB}).  The results as a function of $M_{\rm vir}$ and redshift are shown in Figure \ref{f3a}.  The value of the Peebles spin parameter $\lambda_{\rm P}$ shows a systematic decrease as redshift increases, as was previously noted by \citet{HetzneckerBurkert}, but $\lambda_{\rm B}$ is less dependent on redshift.  Since a smaller value of $\lambda$ is expected to lead to the cooling baryons becoming rotationally supported at a smaller radius, any redshift dependance could have implications for galaxy sizes as a function of redshift.  The latest measurements of galaxy size evolution from HST images \citep{Shibuya+2015,Curtis-Lake+2014} appear to favor the evolution expected from $\lambda_{\rm B}$.

The Tully-Fisher and Faber-Jackson relations relate rotation velocity $V$ and velocity dispersion $\sigma$ of galaxies to their stellar masses. When using 
$V = 1.54\sigma$ and by taking into account the fraction of disk and elliptical galaxies generalize these relations to apply to observed galaxies that have comparable values of $V$ and $\sigma$.  But $\Lambda$CDM simulations show that the cumulative comoving number density of halos with galaxy-scale circular velocities are nearly constant out to rather high redshifts $z\sim 4$ (see Figure \ref{Vmax_cte}), while the stellar mass density decreases with increasing redshift.  This implies that these stellar mass-velocity relations must change at redshifts $z \grtsim 1$.  In Figure \ref{TFfig} we show this, and also show a simple model of how these relations might change up to redshift $z\sim9$.

In Section \ref{LocalVolume} we compare the abundance of dark matter halos with the observed abundance of galaxies in optical and radio surveys.  We show that when we take into account effects of baryons on the observed velocities and the effects of the radii where these velocities are measured, and we compare with the rather wide predicted velocity distribution for volumes of the simulation centered on Milky Way mass galaxies, the agreement between theory and observations is good for low-mass galaxies down to the $\sim 50$ km/s completeness limit of our simulations, contrary to some claims in the literature.

$\Lambda$CDM and observations are also in good agreement at higher masses. 
There are large catalogs of galaxy cluster observations using X-ray and optical surveys, and recently smaller catalogs of cluster Sunyaev-Zel'dovich detections \citep{Planck2015XXIV}.  The key to constructing the cluster mass function from these observations is to obtain a reliable mass calibration from gravitational lensing \citep[e.g.,][]{Rozo+2014,Mantz+2015}.   We have shown that the Tinker mass function, Equation (\ref{Tinker}), is an excellent fit to the abundance of dark matter halos in our simulations, and \citet{Mantz+2015,Planck2015XXIV} find that the predicted and observed cluster abundance are in good agreement with the Planck cosmological parameters.

\section*{Acknowledgments} 
We thank NASA Advanced Supercomputing (NAS) for access to their Pleiades supercomputer where our Bolshoi, MultiDark, and Bolshoi-Planck cosmological simulations were run, with Anatoly Klypin and Joel Primack as leaders.  The MultiDark-Planck simulations were run in the SuperMUC supercomputer at the Leibniz-Rechenzentrum  (LRZ) in Munich, thanks to the computing time provided by the Gauss Centre for Supercomputing  (\url{www.gauss-centre.eu}) and the Partnership for Advanced Supercomputing in Europe (PRACE, \url{www.prace-ri.eu}). Stefan Gottloeber, Anatoly Klypin, Francisco Prada, and Gustavo Yepes were the leaders of the MultiDark-Planck simulation project, and we are very grateful to them for
making the outputs of these simulations available to us.
We also thank the Leibniz Institute for Astrophysics Potsdam (AIP) and the Spanish MultiDark Consolider project for supporting the CosmoSim database (\url{www.cosmosim.org}).  
ARP has been supported by a UC-MEXUS Fellowship. PB was partially supported by a Giacconi Fellowship from the Space Telescope Science Institute (STScI).  The remainder of support for PB through program number HST-HF2-51353.001-A was provided by NASA through a Hubble Fellowship grant from STScI, which is operated by the Association of Universities for Research in Astronomy, Incorporated, under NASA contract NAS5-26555.  AK, CL, and JP acknowledge support from NSF grant AST-1010033  and the  CANDELS grant HST GO-12060 from STScI.  We thank the anonymous Referee for a useful report that improved the presentation of this paper. 
We thank Kristin Riebe for giving us permission to use Fig.~\ref{MergerTrees}, and we thank Vladimir Avila-Reese, Avishai Dekel, Sandra Faber, Stefan Gottloeber, Susan Kassin, David Koo, Gerard Lemson, Francisco Prada, Joe Silk, Rachel Somerville, and Frank van den Bosch for helpful discussions.

\bibliographystyle{mn2efix.bst}
\bibliography{blib,Bolshoi-Planck}

\appendix
\section{Merger Tree Overview}

The publicly available\footnote{\url{http://code.google.com/p/rockstar}} \rockstar\ (Robust Overdensity Calculation using K-Space Topologically Adaptive Refinement) halo finder \citep{ROCKSTAR} identifies dark matter halos based on adaptive hierarchical refinement of friends-of-friends groups of particles in six phase-space dimensions plus time.  For halo masses, \rockstar\ calculates spherical overdensities using all the particles including any substructures in the halo.  Before calculating the halo mass and $V_{\rm max}$, the code performs an unbinding procedure using a modified Barnes-Hut method to accurately calculate particle potentials.
\citet{ROCKSTAR} describes many halo properties that are calculated by \rockstar, and compares results with the \textsc{BDM} halo finder \citep{KlypinGottloeber+1999} that was also used to analyze the Bolshoi simulation \citep{Klypin+2011}.  

\begin{figure}
	\vspace*{-60pt}
	\includegraphics[height=6.5in,width=6.5in]{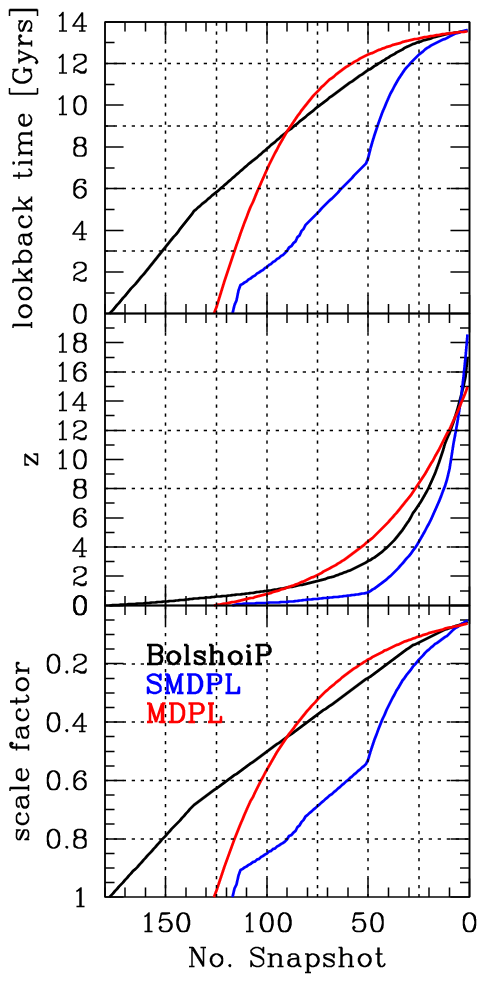}
	\vspace*{-30pt}
		\caption{Lookback time, redshift, and scale factor vs. snapshot number for
		the BolshoiP, SMDPL, and MDPL simulations.  Many high redshift 
		timesteps were saved in order to be able to construct merger 
		trees for halos forming at high redshifts.
 	}
	\label{f2a}
\end{figure}

Results from \textsc{BDM} analyses of all the simulations in Table 1 are available online at the CosmoSim website.$^1$  Those for the 
\rockstar\ analyses of the Bolshoi-Planck, SMDPL, and MDPL simulations can be downloaded in bulk from the UCSC Hyades system.$^2$  Complete particle data was saved for 178 timesteps of the Bolshoi-Planck simulation, 117 timesteps of SMDPL, and 126 timesteps of MDPL.   For scale factor $a=0.06-0.09$ the scale factor change between timesteps was $\Delta a \sim 0.0021$, for $a=0.09-0.14$  $\Delta a \sim 0.0035$,  for $a=0.14-0.7$ $\Delta a \sim 0.005$, and for $a=0.7-1$ $\Delta a = 0.007$.  Thus the time interval between stored timesteps ranges from $\sim 10$ Myr at $z \sim 15$ to $\sim 100$ Myr at $z=0$.  The cadence of these timesteps is shown in Figure \ref{f2a}.

\ctrees\ \citep{CTrees} generates merger trees and halo catalogs in a way that ensures consistency of halo mass, position, and velocity across time steps.  This allows it to repair inconsistencies in halo catalogs---e.g., when a halo disappears for a few time steps, \ctrees\ can regenerate its expected properties by gravitational evolution from the surrounding time steps.
The \ctrees\ code is publicly available,\footnote{\url{https://bitbucket.org/pbehroozi/consistent-trees}} and \ctrees\ outputs used in this paper are available online.$^2$  In these Appendixes we describe how these outputs are organized.

\ctrees{} generates merger tree information in \textbf{tree files} (\texttt{tree\_*.dat}), which each contain halos and their full progenitor histories for cubical subvolumes of the simulation.  It also generates \textbf{catalogs} (\texttt{hlist\_*.list}), which each contain all the halos for a single snapshot of the simulation, along with frequently-used information from their merger histories (e.g., peak mass, half-mass assembly time, mass accretion rate, etc.).  These enable the user to avoid walking the trees for many common applications (e.g., abundance matching).

\subsection{Tree File Layout}

Tree files contain header information (lines beginning with a \texttt{\#} character), a single line with the total number of trees in the file, and then the actual merger trees.  Each merger tree corresponds to the progenitor history of a single halo at the last snapshot of the simulation, containing all halos that fully merge (a.k.a., are disrupted) into the final halo.  Specifically, subhalos that remain distinguishable from their host halo at the last simulation snapshot have separate merger trees in the tree files.  Each tree file contains the merger histories for halos whose centers lie in a cubical subvolume of the simulation.  For example, the 
Bolshoi simulation has 125 (=$5\times{}5\times{}5$) tree files, each corresponding to a $(50$ Mpc/$h)^3$ subvolume of the total $(250$ Mpc/$h)^3$ simulation volume.

The merger trees' basic format is a single header line with the final halo's ID (\texttt{\#tree XYZ}), also known as the \textbf{tree root ID}, followed by single lines for each halo in the full merger history, ordered by snapshot:
\begin{verbatim}
  #tree XYZ
  halo XYZ @ snapshot N
  progenitor 1 @ snapshot N-1
  progenitor 2 @ snapshot N-1
  ...
  progenitor P @ snapshot N-1
  progenitor 1 @ snapshot N-2
  ...
  progenitor Q @ snapshot N-2
  ...
  progenitor 1 @ snapshot 1
  ...
  progenitor Z @ snapshot 1
\end{verbatim}
The format of the halo lines is described in 
Appendix B.
Note here that the order of progenitors within a snapshot is not guaranteed; also, progenitors may not exist at all snapshots, especially if the final halo is close to the mass resolution of the simulation.  Flyby halos (i.e., halos that pass in and out of the virial radius of a host, also called spashback halos) are similarly not guaranteed to be included in the merger history.

For some applications (e.g., semi-analytic models, SAMs), it is necessary to process merger histories including all flyby halos and subhalos.  These are also known as \textbf{forests}.  \ctrees{} provides a \texttt{forests.list} file; each line of the file contains a tree root ID and the corresponding forest ID.  All merger trees with the same forest ID belong to the same forest.  To aid with collating forests from the tree files, \ctrees{} also provides a \texttt{locations.dat} file; each line of the file contains a tree root ID, a tree filename, and the byte offset in the tree file at which the given merger tree may be found.

\subsection{Catalog File Layout}

Catalog files (\texttt{hlist\_*.list}) contain all the halos at a given snapshot of the simulation.  The number in the filename corresponds to the scale factor $a$; e.g., \texttt{hlist\_1.00000.list} corresponds to the simulation output at $a=1.0$.  These files contain header information (lines beginning with a \texttt{\#} character), followed by a single line for each halo in the given snapshot.  The halo line format follows that of the tree files, but includes several additional fields at the end.  All fields are described in 
Appendix B.

\subsection{Alternate Formats}

The above two formats are the only ones generated by default.  Converters also exist for the \textsc{Sussing}, \textsc{Galacticus}, and \textsc{Irate} formats; please contact Peter Behroozi (\texttt{pbehroozi@gmail.com}) for details.

\section{Merger Tree Fields}
\label{s:fields}
Each halo line contains several fields separated by a space character (`` "), giving information about the halo's identification, properties, order within the merger tree, and so on.  Please note that the ordering of some fields may change in new versions of \ctrees{}.  However, the first header line in both the tree files and the catalogs always lists the field order, so it is always a good idea to \textit{double-check this line} with what your code is expecting.

\subsection{Halo Identification and Properties}

\begin{figure*}
	\includegraphics[height=1.7in,width=3in]{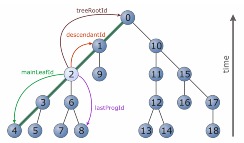}
		\caption{Illustration of a merger tree.  The top node (root) of the tree represents a halo at a given redshift. From there, branches reach backwards in time to its progenitors, i.e., the timeline goes from bottom to top. The numbers at each node indicate the depth-first order, with the most massive progenitors being on the leftmost side of each sub-tree. These form the main branch (e.g., the thick green line for the tree root (0)) of the corresponding node.  (Used with permission from \citet{Riebe}; figure \textcircled{c}2013 WILEY-VCH Verlag GmbH \& Co. KGaA, Weinheim.)}
	\label{MergerTrees}
\end{figure*}

The following fields are always guaranteed to exist, and will always exist in the following order at the beginning of the line:
\begin{itemize}
\item{Scale}: Halo's scale factor.
\item{ID}: Halo ID, guaranteed to be unique across all snapshots of the entire simulation.  Because \ctrees{} adds and deletes halos, this may be different from the ID returned by the halo finder.
\item{Desc\_Scale}: Scale factor of descendent halo, if applicable.
\item{Descid}: Halo ID of descendent halo, if applicable.
\item{Num\_prog}: Number of progenitor halos---i.e., number of halos at the immediately preceding snapshot that fully merge into this halo.
\item{Pid}: Parent halo ID.  For distinct halos (those that are not subhalos), this is \texttt{-1}.  Otherwise, it is the halo ID of the smallest host halo (ordered by $\vmax$) that contains this halo's center within its radius.
\item{Upid}: \"Uber-parent halo ID.  For distinct halos, this is \texttt{-1}.  Otherwise, it is the halo ID of the largest host halo (ordered by $\vmax$) that contains this halo's center within its radius.  Note that it is geometrically possible for a subhalo to be within the virial radius of more than one distinct halo.  Hence, there is no guarantee that the parent halo is a subhalo of the \"uber-parent halo if the two IDs are different.
\item{Phantom}: Zero (i.e., ignorable) for most halos.  This is nonzero only when the halo finder did not return an appropriate progenitor halo at this snapshot and \ctrees{} had to interpolate halo properties from nearby snapshots.
\item{SAM\_Mvir}: \textit{Do not use}.  For historical reasons, this was needed by certain semi-analytical models.
\item{Mvir}: Halo mass, in units of M$_\odot/h$.
\item{Rvir}: Halo radius, in units of comoving kpc$/h$.
\item{Rs}: NFW scale radius, in units of comoving kpc$/h$.
\item{Vrms}: Halo particle velocity dispersion, in units of physical (i.e., non-comoving) km/s.
\item{mmp?}: \texttt{1} if the halo is the most-massive progenitor of its descendent halo; \texttt{0} if not.
\item{scale\_of\_last\_MM}: scale factor of the halo's last major merger.  This is typically defined as a mass ratio greater than 0.3:1, although it is user-adjustable.  The exact definition always appears in the header lines.
\item{Vmax}: Maximum halo circular velocity (i.e., maximum of $\sqrt{GM(<R)/R}$, where $M(<R)$ is the mass enclosed within radius $R$), in units of physical km/s.
\item{X/Y/Z}: Halo position, in units of comoving Mpc/$h$.
\item{VX/VY/VZ}: Halo velocity, in units of physical km/s.
\item{JX/JY/JZ} Halo angular momentum, in units of (M$_\odot/h) \times ($Mpc/$h) \times $ km/s (physical).
\item{Spin}: Dimensionless halo spin parameter.  (Peebles spin, Equation (\ref{lambdaP}), for \textsc{Rockstar}).
\end{itemize}

\subsection{Halo Ordering, Cross-Referencing, and Tree Walking}
When walking trees, there are two main approaches.  The first is \textit{breadth-first}, in which halos are ordered and accessed according to the simulation snapshot; this is the default halo ordering in the trees.  An alternate method is \textit{depth-first}, in which halos are ordered and accessed first along main progenitor branches (i.e., following the most-massive progenitor line) followed by the next-most-massive progenitor branches.  For an illustration, see Figure \ref{MergerTrees}.

The following fields will always exist in current and later versions of \ctrees{}, although more ordering fields may be added at a later date:
\begin{itemize}
\item{Breadth\_first\_ID}: Unique ID (across all simulation snapshots) corresponding to the breadth-first order of halos within a tree.  Sorting on this ID always recovers the original order in which the halos were printed in the tree file.
\item{Depth\_first\_ID}: Unique ID (across all simulation snapshots) corresponding to the depth-first ordering of halos within a tree.  Sorting on this ID reorganizes the tree into depth-first order.  This ordering has the useful property that full merger histories are contiguous, even for halos not at the last simulation snapshot; see Last\_progenitor\_depthfirst\_ID and Last\_mainleaf\_depthfirst\_ID below.
\item{Tree\_root\_ID}: Halo ID of the final descendent halo (i.e., the descendent halo at the last simulation snapshot).
\item{Orig\_halo\_ID}: Original halo ID from halo finder, except for phantom halos.  This allows cross-referencing halos in the merger trees with other data (e.g., particles) saved by the halo finder.
\item{Snap\_num}: Snapshot number from which the halo originated---necessary as not all halo finders generate unique IDs.
\item{Next\_coprogenitor\_depthfirst\_ID}: Depth-first ID of next ``coprogenitor''---i.e., the next halo that shares the same descendent halo.
\item{Last\_progenitor\_depthfirst\_ID}: Depth-first ID of last progenitor.  When a merger tree is sorted by depth-first ID, then the halos between this halo's Depth\_first\_ID and this halo's Last\_progenitor\_depthfirst\_ID correspond to its full merger history.
\item{Last\_mainleaf\_depthfirst\_ID}: Depth-first ID of last progenitor on this halo's main progenitor branch.  When a merger tree is sorted by depth-first ID, then the halos between this halo's Depth\_first\_ID and this halo's Last\_mainleaf\_depthfirst\_ID correspond to its most-massive progenitor history.
\end{itemize}

Note that to better understand the last two fields, it is helpful to refer to Figure \ref{MergerTrees}.

\subsection{Additional Halo Properties}

These fields are halo-finder dependent.  With the current version of \ctrees{} and \textsc{Rockstar}, they include:
\begin{itemize}
\item{Tidal\_Force}: Strongest tidal force from any nearby halo, in dimensionless units ($R_\mathrm{halo} / R_\mathrm{hill}$).   The Hill radius \citep[see, e.g.,][]{Hahn+2009,Hearin+2015} is an upper bound on the spatial extent of newly infalling material that can remain gravitationally bound to a secondary halo.
It is given by $R_\mathrm{hill} = D(M_\mathrm{sec}/3M_\mathrm{prim})^{1/3} = R_\mathrm{sec}(D/3^{1/3} R_\mathrm{prim})$, where the primary (secondary) halo is the larger-virial-radius (smaller-virial-radius) halo of a pair whose centers are separated by distance $D$.  
\item{Tidal\_ID}: Halo ID of halo exerting strongest tidal force.
\item{Rs\_Klypin}: NFW scale radius in units of comoving kpc$/h$, determined using Vmax and Mvir, Equation (\ref{RsKlypin}). 
\item{Mvir\_all}: Halo mass, including unbound particles (M$_\odot/h$).
\item{M200m--M2500c}: Mass (M$_\odot/h$) enclosed within specified overdensities.  These include 200m, 200c, 500c, and 2500c, where $\rho_c$ is critical density and $\rho_m = \Omega_M \rho_c$ is the mean matter density. 
\item{Xoff}: Offset of halo center (defined as the density peak) from the center of mass
within the halo radius (comoving kpc/$h$).  
\item{Voff}: Offset of halo center (defined as the density peak) from average particle velocity (physical km/s).
\item{Spin\_Bullock}: Bullock spin parameter, Equation (\ref{lambdaB}). 
\item{b\_to\_a, c\_to\_a}: Ratio of second and third largest shape ellipsoid axes (B and C) to largest shape ellipsoid axis (A) (dimensionless), calculated according to the method in \citet{Allgood+2006}.  Additional subscripts (e.g., \texttt{500c}) indicate that only particles within a specified halo radius are considered.
\item{A[x],A[y],A[z]}: Largest shape ellipsoid axis (comoving kpc/$h$).  Additional subscripts (e.g., \texttt{500c}) indicate that only particles within a specified halo radius are considered.
\item{T/$|$U$|$}: ratio of kinetic to potential energies for halo particles.
\item{M\_pe\_*}: Pseudo-evolution corrected masses (very experimental).
\end{itemize}

\subsection{Additional Catalog Fields}

These fields are not in the trees, but are provided for convenience in the halo catalogs.  Their order and content may change in future \ctrees{} versions.

\begin{itemize}
\item{$M_\mathrm{acc},V_\mathrm{acc}$}: Halo's (or its main progenitor's) mass and $\vmax$ at the last timestep that it was a distinct halo. 
\item{$M_\mathrm{peak},V_\mathrm{peak}$}: Peak mass and peak $\vmax$ along the halo's main progenitor branch.
\item{Halfmass\_Scale}: Scale factor at which the halo mass in the main progenitor branch first reaches half of Mpeak.
\item{Acc\_Rate\_*}: Halo mass accretion rates in M$_\odot/h$/yr.  These include: \textit{Inst} (averaged since the previous snapshot); \textit{100Myr} (averaged over past 100 Myr); \textit{X*Tdyn}: averaged over past X virial dynamical times (i.e., $X/\sqrt{G\rho_\mathrm{vir}}$); \textit{Mpeak}: averaged growth of Mpeak, averaged from the current halo's redshift ($z$) to $z+0.5$.
\item{Mpeak\_Scale}: Scale at which Mpeak was reached along the main progenitor branch.
\item{Acc\_Scale}: Last scale at which the halo (or its main progenitor) was distinct.
\item{First\_Acc\_Scale}: Last scale at which the halo (or its main progenitor) and all earlier progenitor halos were all distinct.\footnote{In theory.  In practice, flybys often happen for very early progenitors, so these are ignored if halos at least double their mass following the flyby.}
\item{First\_Acc\_(Mvir$|$Vmax)}: Mvir and $\vmax$ at First\_Acc\_Scale for the main progenitor.
\item{Vmax@Mpeak}: Main progenitor's $\vmax$ at Mpeak\_Scale.
\item{Tidal\_Force\_Tdyn}: Dimensionless tidal force averaged over the past virial dynamical time.
\end{itemize}

\end{document}